\documentclass{sigchi}
\usepackage{array}
\usepackage{adjustbox}
\usepackage{multirow}

\toappear{\scriptsize \textbf{[This is a pre-print manuscript]} Permission to make digital or hard copies of all or part of this work for personal or classroom use is granted without fee provided that copies are not made or distributed for profit or commercial advantage and that copies bear this notice and the full citation on the first page. Copyrights for components of this work owned by others than the author(s) must be honored. Abstracting with credit is permitted. To copy otherwise, or republish, to post on servers or to redistribute to lists, requires prior specific permission and/or a fee. Request permissions from permissions@acm.org. \\
{\emph{L@S '21, June 22--25, 2021, Virtual Event, Germany.} } \\
\copyright~2020 Copyright is held by the owner/author(s). Publication rights licensed to ACM. \\
}
\usepackage{balance}       
\usepackage{graphics}      
\usepackage[T1]{fontenc}   
\usepackage{txfonts}
\usepackage{mathptmx}
\usepackage[pdflang={en-US},pdftex]{hyperref}
\usepackage{color}
\usepackage{booktabs}
\usepackage{textcomp}

\usepackage{microtype}        
\usepackage{ccicons}          

\def\plaintitle{Student Barriers to Active Learning in Synchronous Online Classes: Characterization, Reflections, and Suggestions}

\def\emptyauthor{}
\def\plainkeywords{Student-centered education; active learning; synchronous
online classes; interaction; pedagogy.}

\makeatletter
\def\url@leostyle{%
  \@ifundefined{selectfont}{
    \def\UrlFont{\sf}
  }{
    \def\UrlFont{\small\bf\ttfamily}
  }}
\makeatother
\def\pprw{8.5in}
\def\pprh{11in}

\definecolor{linkColor}{RGB}{6,125,233}
\hypersetup{%
  pdftitle={\plaintitle},
  pdfauthor={\emptyauthor},
  pdfkeywords={\plainkeywords},
  pdfdisplaydoctitle=true, 
  bookmarksnumbered,
  pdfstartview={FitH},
  colorlinks,
  citecolor=black,
  filecolor=black,
  linkcolor=black,
  urlcolor=linkColor,
  breaklinks=true,
  hypertexnames=false
}

\begin{document}

\title{\plaintitle}
\numberofauthors{5}
\author{%
  \alignauthor{Reza Hadi Mogavi\\
    \affaddr{HKUST}\\
    \email{rhadimogavi@cse.ust.hk}}\\
  \alignauthor{Yankun Zhao\\
    \affaddr{HKUST}\\
    \email{yzhaock@connect.ust.hk}}\\
  \alignauthor{Ehsan Ul Haq\\
    \affaddr{HKUST}\\
    \email{euhaq@connect.ust.hk}}\\
  \alignauthor{Pan Hui\\
    \affaddr{HKUST \& University of Helsinki}\\
    \email{panhui@cse.ust.hk}}\\
  \alignauthor{Xiaojuan Ma\\
    \affaddr{HKUST}\\
    \email{mxj@cse.ust.hk}}\\
}

\maketitle

\begin{abstract}
As more and more face-to-face classes move to online environments, it becomes increasingly important to explore any emerging barriers to students' learning. This work focuses on characterizing student barriers to active learning in synchronous online environments. The aim is to help novice educators develop a better understanding of those barriers and prepare more student-centered course plans for their active online classes. Towards this end, we adopt a qualitative research approach and study information from different sources: social media content, interviews, and surveys from students and expert educators. Through a thematic analysis, we craft a nuanced list of students' online active learning barriers within the themes of \textit{human-side}, \textit{technological}, and \textit{environmental} barriers. Each barrier is explored from the three aspects of frequency, importance, and exclusiveness to active online classes. Finally, we conduct a summative study with 12 novice educators and explain the benefits of using our barrier list for course planning in active online classes.
\end{abstract}



\section{Introduction}~\label{sec: Introduction}
Over the past decade, online education has been gaining popularity around the world as a way to promote education. Spurred by natural disasters, pandemics, and wars, it is once again gaining more momentum in higher education \cite{10.1145/3313831.3376781}. According to a recent national survey across the United States, more than 80\% of post-secondary institutions are planning to increase their emphasis on developing online programs \cite{morethan80}. 
Online programs can broadly be categorized into synchronous and asynchronous types \cite{10.1145/2702123.2702349, faulconer2018comparison}. In a synchronous class, educators can use applications such as Zoom, Adobe Connect, Minerva, and Blackboard to deliver lectures and carry out activities with their students in real-time \cite{faulconer2018comparison}. 

\begin{figure}[t]
    \centering
    \includegraphics[width=0.40\textwidth]{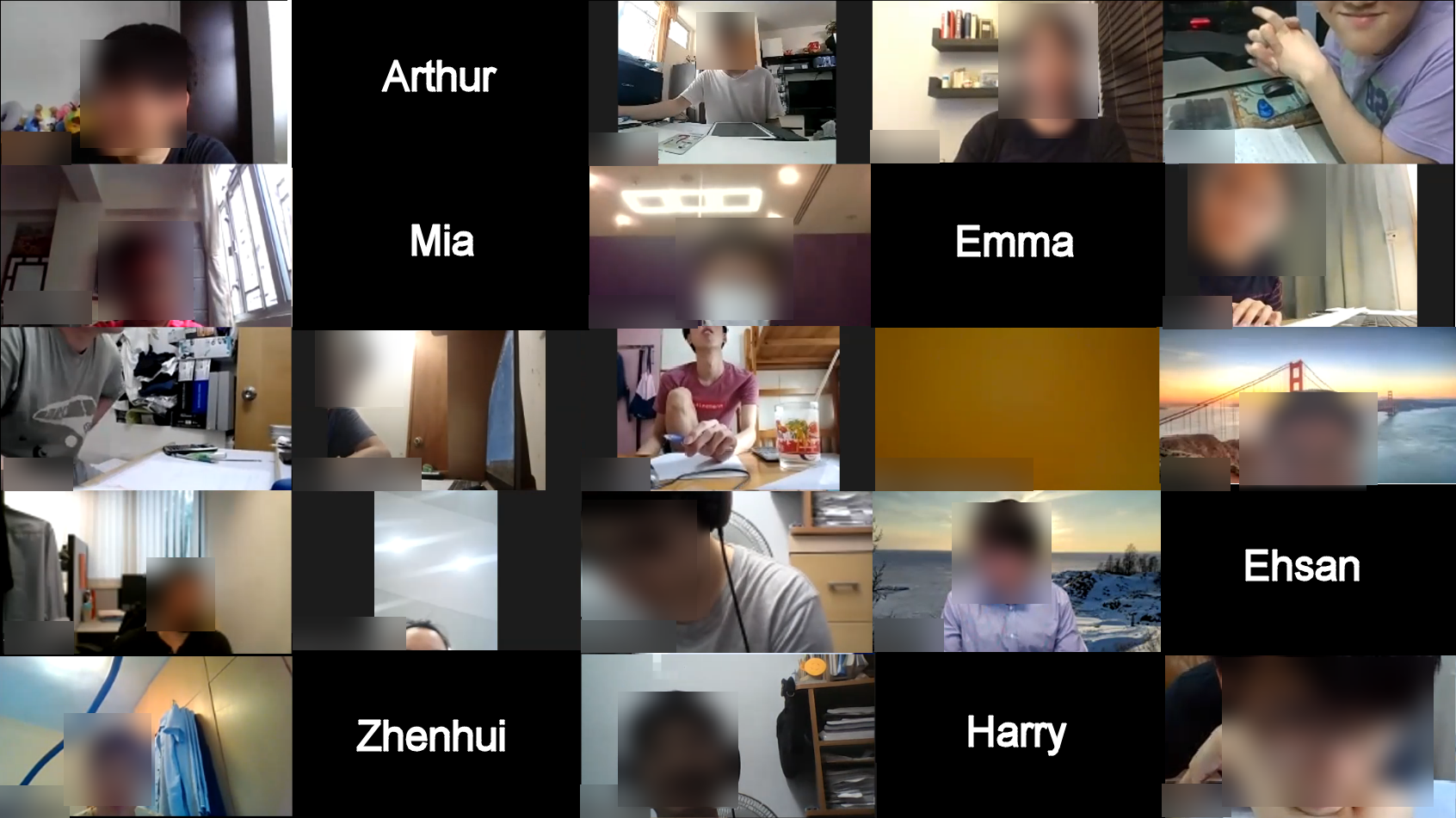}
    \caption{A snapshot from an active online class held through Zoom.}
    \label{fig:synchronous_zoom}
\end{figure}

Despite its popularity, many educational experts argue that an effective synchronous class requires more than just ``\textit{giving every professor a Zoom account and letting instruction take its course} \cite{morethanzoom}.'' Engaging students in vigorously meaningful activities is at the core of success for such classes \cite{10.1145/3290605.3300534}, which demands a well-thought-out teaching methodology (pedagogy) \cite{naeem2020analyzing, weiser2018medium}. Active learning (AL) is one such pedagogy that ``\textit{involves students in doing things and thinking about the things they are doing} \cite{bonwell1991active}.'' Group discussions, simulations, and educational games are among the most familiar AL strategies many educators know and apply in face-to-face classes \cite{abdelsattar2019active}. More than 200 instances of such active learning strategies have been gathered and published online by the University of South Florida \cite{ALStrategies200}. Figure \ref{fig:synchronous_zoom} shows an example of an active online class held synchronously through Zoom.

Nowadays, AL has become an established research track in the fields of Human-Computer Interaction (HCI) and Learning at Scale. Researchers and educators unanimously agree that AL pedagogy is more successful than traditional lecturing methods both for HCI classes and other subjects \cite{10.1145/3313831.3376776, de2019let}. Some of the benefits include but are not limited to students' increased duration of attention and retention in lectures, better critical thinking, improved interpersonal skills, and fewer  course failures \cite{freeman2014active, prince2004does}. Nevertheless, we still hear many educators, especially novices, often fail to achieve their desired AL outcomes \cite{edwards2017like, virtue2018dispositional, cooper2020investigating}.


 This problem seems to be more frequent in synchronous online classes.
 Here, we cite some sample quotes from North America's largest educational development community, called POD Network \cite{PODpeople1} to witness our claim:
\begin{itemize}
    \item [-] ``\textit{As a recently hired adjunct lecturer [in] Greece, me and my students had tons of issues with active online classes, partially because we had \textbf{wrong assumptions and expectations} of one another. ... Teachers: [1] All students like to learn. (wrong!!!) [2] All students can participate [in] classes actively from home. (wrong!!!) }''
    \item [-] ``\textit{As we all realize, inclusive teaching would strive to \textbf{understand the students [on] a finer level} and embrace the diversity in students' learning into the course structure and activity design. But it is hard to do. Remote teaching during the pandemic\footnote{Refers to the COVID-19 pandemic} makes it more challenging.}'' 
\end{itemize}

\noindent \textbf{Research Questions.}
There being a lack of a nuanced view about student barriers to active learning is one problem mentioned by many educators, especially novice ones. Although the literature has already explored students' resistance to active learning in face-to-face classes \cite{tharayil2018strategies, silverthorn2020active}, there is currently a knowledge gap about student barriers in synchronous online classes. 
Our research contributes to this gap by characterizing student barriers to AL in the context of synchronous online classes. 
More formally, this research is guided by the following research questions:

\begin{itemize}
    \item \textbf{RQ1:} What are the \textbf{student barriers} to active learning in synchronous online classes?
    \item \textbf{RQ2:} How do \textbf{expert active learning educators} reflect on the student barriers found?
    \item \textbf{RQ3:} How could \textbf{novice active learning educators} benefit from the student barriers found?
\end{itemize}

Here, we conduct qualitative research following an integrated deductive and inductive approach to answer the questions in three studies. Accordingly, we find and discuss a nuanced list of student barriers within the themes of \textit{human-side}, \textit{technological}, and \textit{environmental} barriers. 

Each study has an extensive research design. Study 1 aims to characterize student barriers by studying their opinions through different sources of \textit{social media content}, \textit{semi-structured interviews}, and \textit{surveys}. Study 2 uses \textit{semi-structured interviews} and \textit{surveys} to explore how expert educators reflect on the barriers found. Finally, study 3 shows the benefits of using our barrier list for planning active online classes by conducting a summative study with 12 novice educators. The summative study is comprised of \textit{surveys} and \textit{follow-up semi-structured interviews}.

\noindent \textbf{Contributions.} This paper makes three main contributions. First, to the best of our knowledge, it is the first comprehensive qualitative research to draw HCI and Learning at Scale researchers' attention to student barriers to active learning in synchronous online classes. The findings can help to inspire new and in-demand research directions for future work. Second, the proposed framework provides a contemporary synthesis of student barriers to online active learning that educators (especially novices) might be less aware of. Finally, our framework can serve as an important construct to explore and research when considering how to create more inclusive and student-centered online classes.

\section{Related Work}\label{sec: Related Work}
\subsection{Synchronous Learning}\label{RW: Synchronous Learning}
\textit{Synchronous learning} means that training is delivered to students in real-time, usually on a set schedule \cite{sharifrazi2019students, 10.1145/3330430.3333653}.
Many researchers believe that synchronous learning can raise student engagement levels and facilitate immediate interaction among students \cite{10.1145/3290605.3300662, 10.1145/2858036.2858184, clark2015comparing}. Educators can use a variety of applications such as Zoom, Adobe Connect, Minerva, and Blackboard to interact with students directly, similar to a face-to-face classroom \cite{huang2020handbook, piskurich2009rapid, phelps2019successful}. 

Educators often consider three features for picking up their digital class applications: alignment with class activities, students' needs, and productivity \cite{10.1145/3290605.3300548}. According to the \textit{Medium Naturalness Theory}, a well-known framework in psychology, the applications that can lead to more natural interactions between teachers and students, similar to face-to-face classes, are better conduits of knowledge sharing \cite{weiser2018medium, kock2005media, kock2011media}. However, novice educators still need to get proper training on how to teach in online synchronous classes if they want to achieve their desired educational outcomes \cite{piskurich2009rapid, 10.1145/3313831.3376418, lindvall2019coherence}. 

Historically, only less than 10\% of online lectures have been delivered synchronously, and it is often considered a relatively more recent online education approach, which has become especially popular after COVID-19 \cite{10.1145/3330430.3333653, phelps2019successful, henriksen2020folk}. Therefore, the literature still needs further exploration of synchronous learning to fully understand its potential, challenges, and opportunities. This research takes the early steps in investigating student barriers to AL in online synchronous classes.    

\subsection{Active Learning}\label{RW: AL}
Research suggests that after 10 to 15 minutes of listening to a lecture, students’ attention often begins to decline significantly \cite{schwartz2019bite, stuart1978medical}. Furthermore, retention of learned information also drops considerably after the first 10 minutes \cite{schwartz2019bite, thomas1972variation}. Such problems can slow down students’ learning, especially when classes last for extended periods \cite{triglianos2017measuring}.

\begin{table*}[t!]
\begin{center}
\caption{Different branches of Active Learning (AL) pedagogy}~\label{tab:AL categories}
\resizebox{\textwidth}{!}{
\begin{tabular}{| m{10em} | m{20.5em} | m{18em} |} 
\hline
\textbf{Category} & \textbf{Description} & \textbf{Example} \\ 
\hline
Problem-based learning & Learning that results from the process of solving an ill-structured, unresolved, or puzzling problem \cite{cattaneo2017telling, pan2020learning, mogavipoolwebsites}. & Medical students are asked to find methods to reduce the side effects of a medication on the elderly.\\
\hline
Discovery-based learning & Learning that results from the process of exploring the environment for discovering concepts or patterns \cite{cattaneo2017telling, 10.1145/2207676.2208358}. & Business students are asked to detect temporal patterns for stock market price changes.\\
\hline
Inquiry-based learning & Learning that results from the process of answering an inquiry (usually a well-defined question) through an inquiry cycle \cite{cattaneo2017telling, 10.1145/3290605.3300774, li2010inquiry}. & Math students are asked to prove the Pythagoras theorem by themselves.\\
\hline
Project-based learning & Learning that results from working on projects over an extended period of time \cite{cattaneo2017telling, 10.1145/3313831.3376650, 10.1145/3290605.3300368}. & Computer science students are asked to design a book shop management system with C\#. \\
\hline
Case-based learning & Learning that results from contextual and real-life case studies \cite{cattaneo2017telling, ouh2019applying}. & Medical students are asked to collaborate with a doctor to examine a real patient.\\
\hline
\end{tabular}
}
\end{center}
\end{table*}

Active learning could help to mitigate such learning problems \cite{10.1145/3313831.3376776, freeman2014active, prince2004does}. AL is a student-centered approach to instruction that is both hands-on and minds-on \cite{mintzes2020constructivism}. The diversity in AL strategies prevents students from getting bored too quickly and keeps them engaged for a longer time \cite{prince2004does, bonwell1991active}. According to \textit{Edgar Dale's cone of learning}, when students get involved in their own learning process by doing activities other than just listening passively, they have a better chance of retaining more than 90\% of what is taught inside a class \cite{goldberg2012active, tantasawat2019attitudes}. 

Active learning stems from the theoretical framework of Constructivism \cite{10.1145/3027063.3053148, 10.1145/3270316.3270611}. According to constructivism, learning is a personal construction that results from an experiential process. From this perspective, researchers have categorized AL pedagogy into five different branches, i.e., \textit{problem-based learning}, \textit{discovery-based learning}, \textit{inquiry-based learning}, \textit{project-based learning}, and \textit{case-based learning} \cite{cattaneo2017telling, mintzes2020active, budhai2017best}. Each category is a broad construct that can be implemented in several different ways (e.g., group works, games, or simulations) \cite{abdelsattar2019active, weston1986selecting}. All of the AL branches could help students acquire and apply content knowledge in practice and develop higher-order thinking skills \cite{yuliati2018influence}. However, each educator might select from them according to his or her own preference, the topic of study, and the students' needs or expectations \cite{bonwell1996active}. We have provided a brief description and an example of each AL branch in Table \ref{tab:AL categories}. 

The literature in the areas of HCI and Learning at Scale is replete with numerous tools for supporting AL in both face-to-face and online classes (e.g., \cite{10.1145/2212776.2212810, 10.1145/3313831.3376776, 10.1145/3313831.3376518, 10.1145/3330430.3333653}). Despite this, however, the utilization of AL in higher education is still underwhelming in practice \cite{harmon2017professional}. Students' resistance and teachers' reluctance to change, administrative roadblocks, and physical settings are the main barriers to applying AL in higher education \cite{tharayil2018strategies, silverthorn2020active}. Nevertheless, these barriers are often too general to be practical for training novice educators \cite{loughran2016teaching}. 

Additionally, student barriers to AL in online classes are mainly left unexplored in the literature \cite{michael2007faculty}. Hence, our research contributes to this knowledge gap and complements the previous works by characterizing student barriers to AL in the context of synchronous online classes in finer detail. By doing so, we also respond to Bernstein’s call for the ``\textit{second generation}'' of research on AL, where researchers are encouraged to add nuance to the literature \cite{cohen2019think}. 

\section{The Studies}
This research comprises three studies, each corresponding to one of our research questions (RQ-RQ3). All of our studies are ethical and approved by the institutional review board (IRB) of our university. In the rest of this section, we provide more information about each study.
\subsection{Study 1: Students' Perceptions}~\label{st1:student}
The first study comprises a qualitative social media content analysis and semi-structured interviews with students, and a final survey collection.
\subsubsection{Social Media Content Analysis} Qualitative content analysis provides a \textit{non-intrusive} way to access and analyze what people really think \cite{kyngas2019application}. Here, we exploit content analysis on some popular social media platforms to find out what students think about AL's barriers in active online classes. The content we analyze comes from five different social media platforms of \textit{Twitter}, \textit{Facebook}, \textit{Reddit}, \textit{Stack Exchange}, and \textit{Google Groups}. We pick these media platforms because they are all well known to the HCI and Learning at Scale research communities and often provide a rich dataset for qualitative analysis.

To find the relevant educational posts and comments in each media, we use \textit{Ubersuggest}, a web-based keyword search engine \cite{UbsuggestEngine}, to find the most popular keywords on the web, which are related to active learning pedagogy. We find 34 keywords with the highest search volume on the web between December 1st, 2019, to April 15th, 2020. These keywords serve as our primary filter to extract the text-based content from each social media platform. 

We either use the Application Programming Interface (API) each social media platform officially provides, or our own customized crawlers based on those APIs to build our qualitative data repository. After this phase, we use natural language processing techniques such as Latent Dirichlet Allocation (LDA) to make our data cleansing procedure easier \cite{guo2017mining}. Finally, we iterate another round of cleaning and manually remove any irrelevant posts and comments that are not related to active learning in online synchronous classes. In total, more than 3,000 posts and comments are analyzed, and we get 77 posts and comments suitable for our content analysis. A significant proportion of posts are filtered out because we have pruned active learning topics from the machine learning community (see \cite{settles2009active}) and contents about the offline classes (see \cite{nicol2018comparison}).

\subsubsection{Semi-structured Interviews}
We use convenience sampling to recruit participants. Recruitment advertisements have been posted on social media and online communities of local universities in Hong Kong. We use an online form to collect volunteers’ demographic information and investigate whether they have taken any active online classes or not. If one has such experiences with AL, we invite him or her to our interview with monetary compensation. All of the participants are asked to sign and return informed consent forms. Eventually, 32 student interviewees (15 females) are recruited from five local universities, whose ages range from 19 to 34 (Mean = 21.7, STD = 4.4). Students come from majors in engineering (N = 16), business (N = 11), science (N = 4), and sociology (N = 1), 84\% of whom are current undergraduate students and 16\% have postgraduate and higher education. 

We send interview questions to the interviewees one day before the interview to help them recall and roughly organize their responses. This is in order to gain as much information as possible from the interviewees' minds and ensure the interview process runs smoothly and efficiently. Each interview is conducted in the interviewee’s preferred language, via the preferred online platform and at the preferred time. Every interviewee is informed and agrees that the interview will be audio-recorded for research purposes, and all the responses are kept confidential and anonymous.

Here, our interview questions are designed based on our previous findings from social media. The structure of our interview questionnaire is that we first inquire about some basic information from the interviewee’s personal experience of synchronous active online learning classes. Then, we ask the interviewee to tell us about any barriers in his or her mind that will hinder student’s participation, engagement, and learning in such active online learning courses. After that, we share some of the barriers previously found in social media with the interviewee as a hint, in order to listen to the interviewee’s ideas or opinions on these barriers. We also encourage our interviewee to critically review his or her experiences with AL again after hearing sample barriers from social media content. This is to ensure they do not leave out any details from their descriptions and confirm their previous responses.

\subsubsection{Survey of Importance and Exclusiveness}
After we crafted the barrier list, to investigate student interviewees’ opinions on the importance and exclusiveness of each barrier, we design a follow-up survey consisting of closed-ended questions using 7-point Likert scales. The survey is distributed to all 32 students, and we get responses from 27 of them (12 females). 

\subsection{Study 2: Expert Educators' Reflections}~\label{st2:exp}
The second study comprises semi-structured interviews with expert educators and a follow-up survey collection.
\subsubsection{Semi-structured Interviews}
In the second study, we take expert educators' opinions about the student barriers found. These educators' rich experience of monitoring and interacting with students through the years could help us to gain a deeper understanding of student barriers and verify our previous findings \cite{10.1145/3313831.3376149}.

Since many educators do not use AL in their online classes and are only accustomed to traditional teacher-centered lecturing, it was more challenging to find the target expert educators. We use a combination of snowball and convenience sampling to recruit expert educators. The participants are among distinguished local educators in Hong Kong who are well-known for using AL strategies in their online classes. We send invitations to those educators and ask respondents to recommend other experienced and interested educators to participate in our research. The inclusion criteria for these educators are: 1) being self-reported experts in conducting active online classes, 2) having experience of at least one such class in the past academic year (i.e., from August 2019 to June 2020), and 3) having more than four years of academic teaching experience (see \cite{ballantyne2007documenting}). 
In total, 8 expert educators (2 females) are recruited from six local universities. Aged between 38 to 61 (Mean $=$ 49.3, STD $=$ 7.9), they are from different departments: engineering (N $=$ 4), science (N $=$ 1), humanities (N $=$ 2), and linguistics (N $=$ 1). Their experience of teaching in AL style ranges from 6 to 25 years (Mean $=$ 13.1, STD $=$ 7.3). 

After taking the expert educators' informed consent, we invite them to participate in an online semi-structured interview using Zoom. Every interview session is audio-recorded for later analysis. During the interview, we ask each participant detailed questions aiming to extract their personal experience about active online classes, their reflections on the barriers found, and the possible solutions.

\subsubsection{Survey of Importance and Exclusiveness.} Since we also want to know our interviewees' opinions about each barrier's importance and exclusiveness to active online classes, we share with each participant a follow-up survey similar to the one we gave to the students. We receive the responses of 7 interviewees (2 females). Only one educator did not respond in a timely fashion. 
\subsection{Study 3: Novice Educators' Evaluation}~\label{st3:novi}
Planning a course is not always straightforward and typically involves a series of steps \cite{weber2020efficient, walker2012course}. \textit{Backward design} is one of the most popular course planning frameworks many educators use \cite{martin2019award, polleck2020putting}. It comprises three stages: 1) defining and prioritizing what students should finally learn in a class, 2) finding suitable assessment measures to make sure students have achieved the intended learning outcomes, and 3) finding the most appropriate instructional activities to carry out in the class. This section conducts a summative study to explore whether and how our barrier list could facilitate novice educators' course planning with backward design.

\subsubsection{Course Planning and Survey Collection}
Again, we use a combination of snowball and convenience sampling for recruiting novice educators. Two postgraduate students disseminate recruitment advertisements on LinkedIn and ask respondents to forward the information to other interested novice educators. The inclusion criteria are being self-reportedly novice educators in conducting active online classes and having less than four years of academic teaching experience (see \cite{ballantyne2007documenting}). Overall, twelve novice educators (3 females) are recruited from six local universities in Hong Kong. They are aged between 32 to 56 (Mean $=$ 40.1, STD $=$ 6.4) and are from different departments: engineering (N $=$ 5), science (N $=$ 3), business and management (N $=$ 2), linguistics (N $=$ 2). Their experience of teaching in AL style ranges from no experience (zero) to 4 years (Mean $=$ 2.1, STD $=$ 1.2).  

After taking the novice educators’ informed consent, we invite them to participate in a summative study to evaluate our barrier list's usefulness and effectiveness. We randomly split participants into two equal groups: \textit{with-list} and \textit{without-list} educators. With-list educators are granted access to check our barrier list, whereas without-list educators are not. We train and ask everyone in both groups to write down a brief course plan for one of the online classes they wish to teach in the AL style. The course topics are the ones these educators should teach in the subsequent semester. Therefore, in this work, our study has a natural setting rather than a synthesized one. 

After consultation with two expert educators we knew from study 2, we give each educator a course planning template that follows backward design steps \cite{bowen2017understanding, Cplanner2017}. The novice educators are given a one-week-long time window to complete and return the template. Each template is accompanied by a survey with some open-ended questions aiming to evaluate novice educators' knowledge about student barriers, the references they use to learn about them, and how they use that information in preparing their course plans. We also ask with-list educators to evaluate our barrier list's usefulness and effectiveness in helping them prepare their course plans. All of our participants return their course plans and survey responses on time.  

\subsubsection{Semi-structured Interviews}
We carefully inspect educators' course plans and responses and summarize our barrier list's benefits (if applicable). Later, after a week, we invite the same participants for a follow-up semi-structured interview. Seven out of 12 participants agree to attend our interviews (i.e., 4 with-list and 3 without-list educators). The interview questions complement the ones from the surveys and give educators a chance to explain their thoughts in finer detail. 

\begin{table*}[t!]
  \centering
  \caption{The triangulated list of student barriers}
 \resizebox{\textwidth}{!}{
        \begin{tabular}{p{9.22em}rrrrrrrrr}
    \toprule
    \multicolumn{1}{c}{} &       &       & \multicolumn{3}{p{11.835em}}{ \centering\textbf{Frequency }} & \multicolumn{2}{p{7.44em}}{\centering\textbf{Importance }} & \multicolumn{2}{p{7.44em}}{\centering\textbf{Exclusiveness }} \\
    \textbf{Barrier Category} & \multicolumn{1}{p{12.055em}}{\textbf{Codes}} & \multicolumn{1}{p{20em}}{\textbf{Sub-codes}} & \multicolumn{3}{p{11.835em}}{\centering\textit{ (Count (\%))}} & \multicolumn{2}{p{7.44em}}{\centering\textit{(Mean (STD))}} & \multicolumn{2}{p{7.44em}}{\centering\textit{(Mean (STD))}} \\
\cmidrule{4-6} \cmidrule{9-10}    \multicolumn{1}{r}{} &       &       & \multicolumn{1}{p{3.945em}}{\textbf{SM (C)}} & \multicolumn{1}{p{3.945em}}{\textbf{ST (I)}} & \multicolumn{1}{p{3.945em}}{\textbf{EXP (I)}} & \multicolumn{1}{p{3.72em}}{\textbf{ST (S)}} & \multicolumn{1}{p{3.72em}}{\textbf{EXP (S)}} & \multicolumn{1}{p{3.72em}}{\textbf{ST (S)}} & \multicolumn{1}{p{3.72em}}{\textbf{EXP (S)}} \\
    \midrule
    \multirow{14}[8]{*}{\textbf{Human-side}} & \multicolumn{1}{l}{\multirow{4}[2]{*}{Affective Barriers [AB] }} & \multicolumn{1}{p{20em}}{[AB 1] Apathy towards active learning} & \multicolumn{1}{p{3.945em}}{\textit{28 (36\%)}} & \multicolumn{1}{p{3.945em}}{\textit{14 (43\%)}} & \multicolumn{1}{p{3.945em}}{\textit{5 (62\%)}} & \multicolumn{1}{p{3.72em}}{\textit{5.2 (0.9)}} & \multicolumn{1}{p{3.72em}}{\textit{4.7 (1.6)}} & \multicolumn{1}{p{3.72em}}{\textit{4.4 (0.8)}} & \multicolumn{1}{p{3.72em}}{\textit{3.8 (1.6)}} \\
    \multicolumn{1}{r}{} &       & \multicolumn{1}{p{20em}}{[AB 2] Low self-efficacy} & \multicolumn{1}{p{3.945em}}{\textit{6 (7\%)}} & \multicolumn{1}{p{3.945em}}{\textit{4 (12\%)}} & \multicolumn{1}{p{3.945em}}{\textit{2 (25\%)}} & \multicolumn{1}{p{3.72em}}{\textit{5.5 (1.0)}} & \multicolumn{1}{p{3.72em}}{\textit{4.8 (1.3)}} & \multicolumn{1}{p{3.72em}}{\textit{4.8 (1.0)}} & \multicolumn{1}{p{3.72em}}{\textit{4.2 (1.3)}} \\
    \multicolumn{1}{r}{} &       & \multicolumn{1}{p{20em}}{[AB 3] Shyness} & \multicolumn{1}{p{3.945em}}{\textit{14 (18\%)}} & \multicolumn{1}{p{3.945em}}{\textit{23 (71\%)}} & \multicolumn{1}{p{3.945em}}{\textit{6 (75\%)}} & \multicolumn{1}{p{3.72em}}{\textit{4.7 (1.2)}} & \multicolumn{1}{p{3.72em}}{\textit{4.7 (1.0)}} & \multicolumn{1}{p{3.72em}}{\textit{3.9 (1.1)}} & \multicolumn{1}{p{3.72em}}{\textit{5.8 (0.6)}} \\
    \multicolumn{1}{r}{} &       & \multicolumn{1}{p{20em}}{[AB 4] Fear of leaving the comfort zone} & \multicolumn{1}{p{3.945em}}{\textit{9 (11\%)}} & \multicolumn{1}{p{3.945em}}{\textit{13 (40\%)}} & \multicolumn{1}{p{3.945em}}{\textit{2 (25\%)}} & \multicolumn{1}{p{3.72em}}{\textit{4.3 (1.2)}} & \multicolumn{1}{p{3.72em}}{\textit{4.1 (1.1)}} & \multicolumn{1}{p{3.72em}}{\textit{3.8 (1.2)}} & \multicolumn{1}{p{3.72em}}{\textit{4.1 (0.6)}} \\
\cmidrule{2-10}    \multicolumn{1}{r}{} & \multicolumn{1}{l}{\multirow{3}[2]{*}{Cognitive Barriers [CB]}} & \multicolumn{1}{p{20em}}{[CB 1] Unbalanced complexity of the active learning activities} & \multicolumn{1}{p{3.945em}}{\textit{7 (9\%)}} & \multicolumn{1}{p{3.945em}}{\textit{6 (18\%)}} & \multicolumn{1}{p{3.945em}}{\textit{3 (37\%)}} & \multicolumn{1}{p{3.72em}}{\textit{4.2 (1.5)}} & \multicolumn{1}{p{3.72em}}{\textit{3.5 (0.9)}} & \multicolumn{1}{p{3.72em}}{\textit{4.8 (1.0)}} & \multicolumn{1}{p{3.72em}}{\textit{4.0 (0.9)}} \\
    \multicolumn{1}{r}{} &       & \multicolumn{1}{p{20em}}{[CB 2] Mental exhaustion} & \multicolumn{1}{p{3.945em}}{\textit{10 (12\%)}} & \multicolumn{1}{p{3.945em}}{\textit{9 (28\%)}} & \multicolumn{1}{p{3.945em}}{\textit{6 (75\%)}} & \multicolumn{1}{p{3.72em}}{\textit{5.1 (1.3)}} & \multicolumn{1}{p{3.72em}}{\textit{3.8 (1.2)}} & \multicolumn{1}{p{3.72em}}{\textit{4.5 (0.9)}} & \multicolumn{1}{p{3.72em}}{\textit{4.3 (0.7)}} \\
    \multicolumn{1}{r}{} &       & \multicolumn{1}{p{20em}}{[CB 3] Absent-mindedness} & \multicolumn{1}{p{3.945em}}{\textit{15 (19\%)}} & \multicolumn{1}{p{3.945em}}{\textit{28 (87\%)}} & \multicolumn{1}{p{3.945em}}{\textit{7 (87\%)}} & \multicolumn{1}{p{3.72em}}{\textit{5.6 (0.8)}} & \multicolumn{1}{p{3.72em}}{\textit{3.8 (1.3)}} & \multicolumn{1}{p{3.72em}}{\textit{5.0 (1.1)}} & \multicolumn{1}{p{3.72em}}{\textit{5.0 (0.5)}} \\
\cmidrule{2-10}    \multicolumn{1}{r}{} & \multicolumn{1}{l}{\multirow{3}[2]{*}{Social Barriers [SB]}} & \multicolumn{1}{p{20em}}{[SB 1] Discrimination} & \multicolumn{1}{p{3.945em}}{\textit{4 (5\%)}} & \multicolumn{1}{p{3.945em}}{\textit{0 (0\%)}} & \multicolumn{1}{p{3.945em}}{\textit{2 (25\%)}} & \multicolumn{1}{p{3.72em}}{\textit{3.4 (2.0)}} & \multicolumn{1}{p{3.72em}}{\textit{2.7 (1.6)}} & \multicolumn{1}{p{3.72em}}{\textit{3.4 (0.9)}} & \multicolumn{1}{p{3.72em}}{\textit{4.0 (0.7)}} \\
    \multicolumn{1}{r}{} &       & \multicolumn{1}{p{20em}}{[SB 2] Harassment} & \multicolumn{1}{p{3.945em}}{\textit{1 (1\%)}} & \multicolumn{1}{p{3.945em}}{\textit{6 (18\%)}} & \multicolumn{1}{p{3.945em}}{\textit{1 (12\%)}} & \multicolumn{1}{p{3.72em}}{\textit{3.6 (1.7)}} & \multicolumn{1}{p{3.72em}}{\textit{3.5 (0.9)}} & \multicolumn{1}{p{3.72em}}{\textit{3.7 (0.8)}} & \multicolumn{1}{p{3.72em}}{\textit{4.8 (1.2)}} \\
    \multicolumn{1}{r}{} &       & \multicolumn{1}{p{20em}}{[SB 3] Isolation} & \multicolumn{1}{p{3.945em}}{\textit{11 (14\%)}} & \multicolumn{1}{p{3.945em}}{\textit{27 (84\%)}} & \multicolumn{1}{p{3.945em}}{\textit{4 (50\%)}} & \multicolumn{1}{p{3.72em}}{\textit{4.8 (1.4)}} & \multicolumn{1}{p{3.72em}}{\textit{4.7 (1.3)}} & \multicolumn{1}{p{3.72em}}{\textit{5.3 (1.3)}} & \multicolumn{1}{p{3.72em}}{\textit{4.6 (1.0)}} \\
\cmidrule{2-10}    \multicolumn{1}{r}{} & \multicolumn{1}{l}{\multirow{4}[2]{*}{Teaching Barriers [TB]}} & \multicolumn{1}{p{20em}}{[TB 1] Poor interaction management} & \multicolumn{1}{p{3.945em}}{\textit{8 (10\%)}} & \multicolumn{1}{p{3.945em}}{\textit{26 (81\%)}} & \multicolumn{1}{p{3.945em}}{\textit{7 (87\%)}} & \multicolumn{1}{p{3.72em}}{\textit{5.1 (1.1)}} & \multicolumn{1}{p{3.72em}}{\textit{4.3 (0.5)}} & \multicolumn{1}{p{3.72em}}{\textit{5.1 (1.3)}} & \multicolumn{1}{p{3.72em}}{\textit{5.0 (0.5)}} \\
    \multicolumn{1}{r}{} &       & \multicolumn{1}{p{20em}}{[TB 2] The educator’s pressure to participate} & \multicolumn{1}{p{3.945em}}{\textit{6 (7\%)}} & \multicolumn{1}{p{3.945em}}{\textit{6 (18\%)}} & \multicolumn{1}{p{3.945em}}{\textit{3 (37\%)}} & \multicolumn{1}{p{3.72em}}{\textit{4.6 (1.5)}} & \multicolumn{1}{p{3.72em}}{\textit{4.6 (0.7)}} & \multicolumn{1}{p{3.72em}}{\textit{4.2 (1.3)}} & \multicolumn{1}{p{3.72em}}{\textit{4.2 (0.4)}} \\
    \multicolumn{1}{r}{} &       & \multicolumn{1}{p{20em}}{[TB 3] Unfair evaluation measures} & \multicolumn{1}{p{3.945em}}{\textit{2 (2\%)}} & \multicolumn{1}{p{3.945em}}{\textit{17 (53\%)}} & \multicolumn{1}{p{3.945em}}{\textit{2 (25\%)}} & \multicolumn{1}{p{3.72em}}{\textit{5.1 (1.2)}} & \multicolumn{1}{p{3.72em}}{\textit{4.1 (0.8)}} & \multicolumn{1}{p{3.72em}}{\textit{4.4 (0.7)}} & \multicolumn{1}{p{3.72em}}{\textit{5.5 (0.7)}} \\
    \multicolumn{1}{r}{} &       & \multicolumn{1}{p{20em}}{[TB 4] Heavy workload} & \multicolumn{1}{p{3.945em}}{\textit{16 (20\%)}} & \multicolumn{1}{p{3.945em}}{\textit{14 (43\%)}} & \multicolumn{1}{p{3.945em}}{\textit{4 (50\%)}} & \multicolumn{1}{p{3.72em}}{\textit{4.5 (1.4)}} & \multicolumn{1}{p{3.72em}}{\textit{4.0 (0.9)}} & \multicolumn{1}{p{3.72em}}{\textit{4.1 (0.7)}} & \multicolumn{1}{p{3.72em}}{\textit{4.8 (0.6)}} \\
    \midrule
    \multirow{8}[6]{*}{\textbf{Technological}} & \multicolumn{1}{l}{\multirow{6}[2]{*}{UX Barriers [UB]}} & \multicolumn{1}{p{20em}}{[UB 1] Inadequate technology support for active learning activities} & \multicolumn{1}{p{3.945em}}{\textit{7 (9\%)}} & \multicolumn{1}{p{3.945em}}{\textit{17 (53\%)}} & \multicolumn{1}{p{3.945em}}{\textit{5 (62\%)}} & \multicolumn{1}{p{3.72em}}{\textit{4.5 (1.5)}} & \multicolumn{1}{p{3.72em}}{\textit{4.1 (0.9)}} & \multicolumn{1}{p{3.72em}}{\textit{5.5 (0.8)}} & \multicolumn{1}{p{3.72em}}{\textit{4.7 (0.7)}} \\
    \multicolumn{1}{r}{} &       & \multicolumn{1}{p{20em}}{[UB 2] Lack of technological support for backchanneling} & \multicolumn{1}{p{3.945em}}{\textit{12 (15\%)}} & \multicolumn{1}{p{3.945em}}{\textit{8 (25\%)}} & \multicolumn{1}{p{3.945em}}{\textit{1 (12\%)}} & \multicolumn{1}{p{3.72em}}{\textit{4.8 (1.5)}} & \multicolumn{1}{p{3.72em}}{\textit{3.5 (1.7)}} & \multicolumn{1}{p{3.72em}}{\textit{5.9 (1.0)}} & \multicolumn{1}{p{3.72em}}{\textit{6.5 (0.5)}} \\
    \multicolumn{1}{r}{} &       & \multicolumn{1}{p{20em}}{[UB 3] Unfriendly interface design} & \multicolumn{1}{p{3.945em}}{\textit{3 (3\%)}} & \multicolumn{1}{p{3.945em}}{\textit{6 (18\%)}} & \multicolumn{1}{p{3.945em}}{\textit{2 (25\%)}} & \multicolumn{1}{p{3.72em}}{\textit{4.2 (1.7)}} & \multicolumn{1}{p{3.72em}}{\textit{3.8 (1.2)}} & \multicolumn{1}{p{3.72em}}{\textit{5.6 (1.0)}} & \multicolumn{1}{p{3.72em}}{\textit{4.7 (0.7)}} \\
    \multicolumn{1}{r}{} &       & \multicolumn{1}{p{20em}}{[UB 4] Poor privacy protection} & \multicolumn{1}{p{3.945em}}{\textit{21 (27\%)}} & \multicolumn{1}{p{3.945em}}{\textit{10 (31\%)}} & \multicolumn{1}{p{3.945em}}{\textit{8 (100\%)}} & \multicolumn{1}{p{3.72em}}{\textit{3.9 (1.8)}} & \multicolumn{1}{p{3.72em}}{\textit{5.3 (0.9)}} & \multicolumn{1}{p{3.72em}}{\textit{5.3 (1.2)}} & \multicolumn{1}{p{3.72em}}{\textit{6.2 (0.8)}} \\
    \multicolumn{1}{r}{} &       & \multicolumn{1}{p{20em}}{[UB 5] Poor accessibility} & \multicolumn{1}{p{3.945em}}{\textit{2 (2\%)}} & \multicolumn{1}{p{3.945em}}{\textit{0 (0\%)}} & \multicolumn{1}{p{3.945em}}{\textit{1 (12\%)}} & \multicolumn{1}{p{3.72em}}{\textit{4.5 (1.8)}} & \multicolumn{1}{p{3.72em}}{\textit{4.1 (1.5)}} & \multicolumn{1}{p{3.72em}}{\textit{4.9 (1.0)}} & \multicolumn{1}{p{3.72em}}{\textit{6.6 (0.5)}} \\
    \multicolumn{1}{r}{} &       & \multicolumn{1}{p{20em}}{[UB 6] Glitches and errors} & \multicolumn{1}{p{3.945em}}{\textit{4 (5\%)}} & \multicolumn{1}{p{3.945em}}{\textit{5 (15\%)}} & \multicolumn{1}{p{3.945em}}{\textit{2 (25\%)}} & \multicolumn{1}{p{3.72em}}{\textit{4.8 (1.7)}} & \multicolumn{1}{p{3.72em}}{\textit{5.2 (0.8)}} & \multicolumn{1}{p{3.72em}}{\textit{5.3 (1.1)}} & \multicolumn{1}{p{3.72em}}{\textit{6.2 (0.7)}} \\
\cmidrule{2-10}    \multicolumn{1}{r}{} & \multicolumn{1}{p{12.055em}}{Network Barrier [NB]} & \multicolumn{1}{p{20em}}{[NB 1] Low internet bandwidth} & \multicolumn{1}{p{3.945em}}{\textit{20 (25\%)}} & \multicolumn{1}{p{3.945em}}{\textit{30 (93\%)}} & \multicolumn{1}{p{3.945em}}{\textit{6 (75\%)}} & \multicolumn{1}{p{3.72em}}{\textit{5.3 (1.7)}} & \multicolumn{1}{p{3.72em}}{\textit{5.8 (0.6)}} & \multicolumn{1}{p{3.72em}}{\textit{6.2 (0.9)}} & \multicolumn{1}{p{3.72em}}{\textit{5.5 (0.5)}} \\
\cmidrule{2-10}    \multicolumn{1}{r}{} & \multicolumn{1}{p{12.055em}}{Economic Barrier [EB]} & \multicolumn{1}{p{20em}}{[EB 1] The high cost of using technologies} & \multicolumn{1}{p{3.945em}}{\textit{6 (7\%)}} & \multicolumn{1}{p{3.945em}}{\textit{6 (18\%)}} & \multicolumn{1}{p{3.945em}}{\textit{5 (62\%)}} & \multicolumn{1}{p{3.72em}}{\textit{4.0 (1.6)}} & \multicolumn{1}{p{3.72em}}{\textit{4.1 (1.5)}} & \multicolumn{1}{p{3.72em}}{\textit{5.5 (1.0)}} & \multicolumn{1}{p{3.72em}}{\textit{6.0 (0.7)}} \\
    \midrule
    \multirow{6}[4]{*}{\textbf{Environmental}} & \multicolumn{1}{l}{\multirow{2}[2]{*}{Geographical Barriers [GB]}} & \multicolumn{1}{p{20em}}{[GB 1] Time zone conflicts} & \multicolumn{1}{p{3.945em}}{\textit{9 (11\%)}} & \multicolumn{1}{p{3.945em}}{\textit{19 (59\%)}} & \multicolumn{1}{p{3.945em}}{\textit{5 (62\%)}} & \multicolumn{1}{p{3.72em}}{\textit{4.4 (1.5)}} & \multicolumn{1}{p{3.72em}}{\textit{3.8 (0.9)}} & \multicolumn{1}{p{3.72em}}{\textit{5.8 (1.1)}} & \multicolumn{1}{p{3.72em}}{\textit{6.2 (0.4)}} \\
    \multicolumn{1}{r}{} &       & \multicolumn{1}{p{20em}}{[GB 2] Regional internet censorship} & \multicolumn{1}{p{3.945em}}{\textit{13 (16\%)}} & \multicolumn{1}{p{3.945em}}{\textit{11 (34\%)}} & \multicolumn{1}{p{3.945em}}{\textit{5 (62\%)}} & \multicolumn{1}{p{3.72em}}{\textit{4.8 (1.6)}} & \multicolumn{1}{p{3.72em}}{\textit{4.8 (1.2)}} & \multicolumn{1}{p{3.72em}}{\textit{6.0 (1.1)}} & \multicolumn{1}{p{3.72em}}{\textit{6.5 (0.5)}} \\
\cmidrule{2-10}    \multicolumn{1}{r}{} & \multicolumn{1}{l}{\multirow{4}[2]{*}{Workspace Barriers [WB]}} & \multicolumn{1}{p{20em}}{[WB  1] Insufficient workspace size} & \multicolumn{1}{p{3.945em}}{\textit{10 (12\%)}} & \multicolumn{1}{p{3.945em}}{\textit{14 (43\%)}} & \multicolumn{1}{p{3.945em}}{\textit{8 (100\%)}} & \multicolumn{1}{p{3.72em}}{\textit{4.5 (1.5)}} & \multicolumn{1}{p{3.72em}}{\textit{4.2 (1.9)}} & \multicolumn{1}{p{3.72em}}{\textit{5.2 (1.3)}} & \multicolumn{1}{p{3.72em}}{\textit{6.1 (0.6)}} \\
    \multicolumn{1}{r}{} &       & \multicolumn{1}{p{20em}}{[WB 2] Unavailability of utilities and tools} & \multicolumn{1}{p{3.945em}}{\textit{8 (10\%)}} & \multicolumn{1}{p{3.945em}}{\textit{15 (46\%)}} & \multicolumn{1}{p{3.945em}}{\textit{7 (87\%)}} & \multicolumn{1}{p{3.72em}}{\textit{4.4 (1.6)}} & \multicolumn{1}{p{3.72em}}{\textit{4.5 (1.5)}} & \multicolumn{1}{p{3.72em}}{\textit{5.5 (1.0)}} & \multicolumn{1}{p{3.72em}}{\textit{5.7 (0.8)}} \\
    \multicolumn{1}{r}{} &       & \multicolumn{1}{p{20em}}{[WB 3] Noise and overcrowding} & \multicolumn{1}{p{3.945em}}{\textit{14 (18\%)}} & \multicolumn{1}{p{3.945em}}{\textit{11 (34\%)}} & \multicolumn{1}{p{3.945em}}{\textit{8 (100\%)}} & \multicolumn{1}{p{3.72em}}{\textit{4.5 (1.7)}} & \multicolumn{1}{p{3.72em}}{\textit{4.3 (1.4)}} & \multicolumn{1}{p{3.72em}}{\textit{4.7 (1.4)}} & \multicolumn{1}{p{3.72em}}{\textit{6.7 (0.4)}} \\
    \multicolumn{1}{r}{} &       & \multicolumn{1}{p{20em}}{[WB 4] Conflicting with home responsibilities} & \multicolumn{1}{p{3.945em}}{\textit{3 (3\%)}} & \multicolumn{1}{p{3.945em}}{\textit{4 (12\%)}} & \multicolumn{1}{p{3.945em}}{\textit{6 (75\%)}} & \multicolumn{1}{p{3.72em}}{\textit{4.1 (1.6)}} & \multicolumn{1}{p{3.72em}}{\textit{4.2 (1.0)}} & \multicolumn{1}{p{3.72em}}{\textit{5.4 (1.1)}} & \multicolumn{1}{p{3.72em}}{\textit{5.6 (0.9)}} \\
    \midrule
    \multicolumn{10}{p{67.99em}}{\textbf{SM $=$ Social Media, ST $=$ Students, EXP $=$ Expert Educators / C $=$ Content Analysis, I $=$ Semi-structured Interviews, S $=$ Surveys}} \\
    \bottomrule
    \end{tabular}}%
  \label{tab: CodeBook}%
\end{table*}%
\section{Findings}
Our findings are presented in four parts: 1) description of the barriers, 2) their frequencies, importance, and exclusiveness, 3) main problems, and 4) benefits of our findings for novice educators. 
\subsection{Description of the Barriers}
Table \ref{tab: CodeBook} provides a summary view of the characterized student barriers from study 1 and 2. The barriers are within the themes of \textbf{human-side} (4 codes, 14 sub-codes), \textbf{technological} (3 codes, 8 sub-codes), and \textbf{environmental} (2 codes, 6 sub-codes) barriers. Each barrier is triangulated using social media content and interviews with students and expert educators. 

$\bullet$ \textbf{Human-side barriers.} These barriers can be categorized into four types of \textbf{affective}, \textbf{cognitive}, \textbf{social}, and \textbf{teaching} barriers. We are inspired by the famous \textit{educational framework of COI} (Community Of Inquiry) for choosing the names of these codes \cite{garrison1999critical}. 

Affective barriers are the ones that damage students' psychological, emotional and social well-being. The affective barriers we find include apathy towards AL, low self-efficacy (not believing in or underestimating one's own capabilities \cite{ghergulescu2019interactive}), shyness, and fear of leaving one's comfort zone (see \cite{aksit2016active}). We find that since most students have got used to ordinary online classes without any AL activities, unfamiliarity and uncertainty usually tend to make them act restrained when they jump into the AL classes.

Cognitive barriers are the ones that stagnate students' thinking and processing of information. The cognitive barriers found include an unbalanced complexity of the AL activities (strategies), mental-exhaustion, and absent-mindedness (see \cite{hollis2016mind}). Our findings imply that the AL strategies students prefer are the ones that put them in a mental state called \textit{flow zone} \cite{mogavi2019hrcr, csikszentmihalyi2020finding}. 
It means that AL strategies should be neither too complex nor too dull or boring. Students also mention social media platforms, video broadcasting websites (e.g., YouTube), and online games to be among the main causes of their absent-mindedness inside active online classes. This is interesting because it seems that active online classes have inherited such general barriers from online classes despite their attempt to engage students more actively. Our findings suggest that one possible reason is that educators have generally less authority in active online classes.

Social barriers include discrimination, harassment, and isolation. According to our observations, discrimination and harassment are among the least mentioned barriers, but they do exist. 
Therefore, these barriers demand more subtle attention from both students and educators. 

Teaching barriers students have experienced include poor interaction management between students and educators, the educators' pressure to participate in AL activities (sometimes with a particular requirement like keeping the cameras on), unfair evaluation measures (or grading style), and heavy workload.

$\bullet$ \textbf{Technological barriers.} These barriers can break down into three categories of user experience (UX) \textbf{design issues}, \textbf{network}, and \textbf{economic} barriers. The instances of UX design barriers include inadequate technological support for AL activities and lack of technological support for backchanneling (see \cite{preece2015interaction}), unfriendly interface design, inadequate privacy protection, poor accessibility for students with special needs (see \cite{potluri2018codetalk}), and glitches and errors in the broadcasting systems or other applications used. Furthermore, network barriers are mostly related to low internet bandwidth, and economic barriers are mostly related to the high cost of using technologies in AL classes. 
 
$\bullet$ \textbf{Environmental barriers.} They are either \textbf{geographical} or \textbf{workspace} related. Geographical barriers include time zone conflicts and regional internet censorship. Since students attend active online classes from all around the globe, their local time zones are more likely to conflict with class hours. Furthermore, regional internet censorship might block some students' access to some learning applications or platforms. Workspace barriers include insufficient workspace at home, unavailability of utilities and tools, noise and overcrowding, and schedule conflicts which can happen due to unplanned or sudden home responsibilities.  

\subsection{Frequencies, Importance, and Exclusiveness}
Table \ref{tab: CodeBook} also shows the frequency with which each barrier was mentioned from different sources. Furthermore, each barrier's importance and exclusiveness to active online classes are also provided according to students' and expert educators' survey responses.

Frequencies show the number of times data from different triangulation sources (social media content, interviews with students, or expert educators) indicate a particular barrier. In total, we have analyzed the content of social media posts from the popular social media platforms of Twitter (N $=$ 10), Facebook (N $=$ 24), Reddit (N $=$ 21), Stack Exchange (N $=$ 4), and Google Groups (N $=$ 18). We have also interviewed 32 students and 8 expert AL educators from local universities. According to Table \ref{tab: CodeBook}, our findings show that the themes of human-side (\textit{total count $=$ 384}), technological (\textit{total count $=$ 187}), and environmental barriers (\textit{total count $=$ 170}) sequentially have the highest frequency of the barriers mentioned by different sources. 
Respectively, affective barriers (\textit{total count $=$ 126}) and economic barriers (\textit{total count $=$ 17}) hold the highest and lowest frequencies among all of the codes. 

We ask the student interviewees (\textit{N $=$ 27}) and expert educators (\textit{N $=$ 7}) to indicate, based on a 7-point Likert Scale, what they think about each barrier's importance (\textit{1: Not important at all, 4: Neutral, 7: Extremely important}) and its exclusiveness to active online classes (\textit{1: Face-to-Face classes only, 4: Neutral, 7: Synchronous online classes only}). Here, we report the mean and standard deviation (STD) for the participants' answers. 

According to the students' responses, absent-mindedness, low self-efficacy, and low internet bandwidth are the most significant barriers to active learning in online synchronous classes. Furthermore, they think that low internet bandwidth, regional internet censorship, and lack of technological support for backchanneling are the most exclusive barriers to active online classes. 

Besides that, expert educators think that technological barriers are the most important barriers. They consider low internet bandwidth, poor privacy protection, and glitches and errors as the most significant barriers their students face. They have prioritized noise and overcrowding, poor accessibility, and regional internet censorship as more exclusive problems to active online classes. Surprisingly, poor accessibility is more exclusive to active online classes than face-to-face classes. One possible explanation we infer based on our other qualitative data is that it is more difficult for educators to inspect students' special needs when teaching online. Educators think that such problems could get even more complicated when students try to hide their disabilities or other accessibility needs.

In conclusion, it is noted that educators’ mindset about a barrier might significantly differ from students’ mindset regarding the same barrier in terms of frequency, importance, and exclusiveness to AL online classes. Narrowing down such perceptive gaps can help educators improve their AL classes' quality and lead to better student learning. 

\subsection{What are the Main Problems?}
Based on the first two studies, we know that there are many barriers to active learning in synchronous online classes, and it is impractical to solve all the 28 granular barriers listed in our list at one time. Therefore, we have summarized 6 main problems of relatively high importance that are particularly worthy of attention by HCI and Learning at Scale researchers and practitioners.

\noindent\textbf{1. Staying focused is still the biggest challenge for students in active online classes.}
As mentioned in nearly ninety percent of students' survey responses, absent-mindedness is regarded to be the most significant barrier in synchronous online classes. Also, students’ opinions on this barrier show the highest degree of consistency among all the 28 barriers. Students indicate that there are many more distractions in online environments and their workspaces, which disengage them and allow their minds to wander. Many students admit that they chat online, browse social media, and watch YouTube videos more frequently with the absence of the educator’s supervision and classroom atmosphere. They also say that it is so easy for them to find distractions like a comfortable bed, delicious snacks, and even video game consoles. Besides that, about one-third of the student interviewees mention that active online classes are more fatiguing than face-to-face active classes, and thus prevent them from staying focused for long periods of time. Physically, staring at screens for a long time makes their eyes very tired. Psychologically, students argue that they have to try to be more intent during such active online classes to avoid missing any information since it is almost impossible for them to get any whispered hints from their classmates. At the same time, they also need to resist all distractions from their workspace. 
\begin{itemize}
    
    \item [-] ``\textit{There's no one to watch you if you're paying attention or not. Because in (physical) class you would feel bad if everyone is taking notes and everyone is listening, and you just play Instagram, it would be really weird. But then actually you are just in your room, and the video is not even on. It's really easy to be distracted. Sometimes I would even just put my lecture in the background noise and I will do something else so that I don't feel guilty for missing the class... And I know a lot of my friends also do this.}'' (Undergraduate, Finance \& Information System)
    
\end{itemize}

\noindent\textbf{2. Incentives for online active learning might be insufficient and unfair.} 
Students' responses in the interview show that most students are neutral during online active learning. They do not have much enthusiasm for this learning strategy, and just want to follow the educator’s instructions and complete tasks, in order to get good grades. However, as many student interviewees mention, incentives (mainly refer to grade incentives, like participation points) are too minor to keep them engaged in their active online classes. Besides that, sometimes the unfairness and opacity of the grading scheme also confuses students and decreases their motivation.
\begin{itemize}
    \item [-] ``\textit{Well, after all, if there are no grade incentives, in fact, most people may not be willing to speak in this kind of online class. […] My overall feeling is that, if the instructor does not state that your participation would be counted towards your grade, then most students won’t participate a lot in such [an] online setting. But in face-to-face classes, even though your speaking may not be scored, I think some people will still raise their hands.}'' (Undergraduate, Decision Analytics)
\end{itemize}

\noindent\textbf{3. Students cannot get timely help from the teacher.}
About eighty percent of student interviewees complain that the interaction in their active online classes is poorly managed by the educator. They generally attribute this issue to educators and technology. Some students think educators need more training to become proficient in online education software, but sometimes there are indeed too many messages to handle, and others feel that the interaction functions of the online education software they use are not well designed. 
\begin{itemize}
    \item [-] ``\textit{For the (online class) meeting, there was only one instructor, and there were a lot of students. And some of us could raise questions into the chat box. So, he probably couldn't manage all of them, like seeing the raised hands and the chat box at the same time, and he also needs to deliver the lecture at the same time. I think it’s kind of difficult for the professor as well.}'' (Undergraduate, Food and Nutritional Sciences)
    \item [-] ``\textit{If in some face-to-face group discussions, when the teacher poses a question to a group, they may have no idea about it at first, then they can easily ask the teacher for help. […] However, in online group discussions, it will be difficult for them to do this. If they are already in a breakout room for group discussion, then they are literally isolated. If the teacher doesn't enter their breakout room, it is very hard and inconvenient for them to get any help. The group discussion will probably get stuck if they meet any problem.}'' (Undergraduate, Electronic Engineering)
\end{itemize}

\noindent\textbf{4. The communication efficiency of group work in an online setting is lower than face-to-face.}
Over eighty percent of student interviewees express their feelings of isolation. They feel depressed and lonely because they cannot talk to their classmates face-to-face, which hinders them from making friends. Such forced estrangement reduces the efficiency of idea exchange in their online active learning activities such as group projects. Additionally, backchannels always facilitate human communication, which means in a group discussion, the listeners usually use brief vocalizations (e.g., ``yeah'' and ``uh-huh''), gestures, facial expressions, and eye contact to indicate their attention to, understanding of, or agreement with the speaker. Students agree that the loss of such backchannels in online communication usually embarrasses the speaker and dulls the discussion.
\begin{itemize}
    \item [-] ``\textit{Since it’s the online class, in most of the classes, I do not know anyone. So, everyone is new to me. I think it's really hard to make friends during online classes because we cannot meet them physically. Even though we have any group work, we only discuss about the group work. And we don't talk about like anything else, and we do not become friends. So, I think that is a big problem for me, too. […] Since we are not familiar with each other, it's really difficult for us (to communicate efficiently), like, since there are discussions, I think there are a lot of significant opinions from my classmates. But if we don't really know each other, I think it's hard for us to understand what they are trying to say.}'' (Undergraduate, Professional Accountancy)
    \item [-] ``\textit{Several times when I was entering my language course, I was assigned into a breakout room. And then I find out no one is turning on the camera. […] It’s a bit embarrassing because when I say something virtually, I don't know whether they have responses or not. If you do this physically, you can meet them face to face. Then you can talk to them. You can look at their eyes and they will respond to you. But in the virtual world, I don't know whether they will respond to me. So, sometimes I don't really want to talk virtually, because I just feel awkward and it really impedes the active participation part.}'' (Undergraduate, Computer Science)
\end{itemize}

\noindent\textbf{5. Active learning could be even more time-consuming online.}
AL activities are often criticized for being too time-consuming \cite{michael2007faculty}. More than half of the expert educators we interviewed believe that this problem has worsened in active online classes. One educator mentions that a ten-minute AL activity in a face-to-face class could easily take up to half an hour in an online version. The common reasons expert educators point out include poor backchanneling and students' lack of familiarity with how the technology works. 
\begin{itemize}
    \item [-] ``\textit{It seems to be a common phenomenon in online environments. When we talk to black screens (people with their cameras turned off), we repeat ourselves more than before to make sure even those we cannot see are understanding. But I think if everyone [turns on] their cameras, this will become less issue. [...] Our body language help to adjust the pace of our knowledge transfer. If we guess, we get naturally more delayed.}'' (Professor, Computer Science and Engineering) 
\end{itemize}

\noindent\textbf{6. Active online classes are more prone to be inequitable.}
Many expert educators we interviewed believe that reaching equity in active online classes is more challenging than face-to-face classes. Compared to face-to-face classes, it is more difficult for educators to discover barriers students are experiencing, because they have less contact with students. For example, not all of the students have equal socio-economical statuses, which could limit their access and usage of needed technologies and tools. Many student interviewees report that they do not have stable utility services (e.g., internet and electricity) or sufficient private workspace, and they are not equipped with adequate devices for active online classes such as powerful computers, cameras, and microphones.
\begin{itemize}
    \item [-] ``\textit{I think the most common one is the technological barrier. For example, WIFI connections, speaker function, camera function. I think those technological barriers are the most common one because not everyone is fortunate enough to have a strong WIFI and a very good computer.}'' (Undergraduate, Finance)
\end{itemize}
Furthermore, some expert educators mention that interaction between them and students can also be problematic. For example, Zoom, a widely used platform for online classes, can only display 25 students at a time on one screen, and educators could unintendedly interact only with those 25 students and neglect the others. Such inequity is particularly pronounced in large-scale classes. Additionally, students with their cameras on or off may also create some bias in educators' perceptions of those students.
\begin{itemize}
    \item [-] ``\textit{At a face-to-face class, teachers can check from the front line of class to the very end in just one sight. At Zoom, I can see only 25 people in one sight. It is more time-taking to interact with a large class and some students may receive less attention.}'' (Associate Professor, Computer Science and Engineering)
    \item [-] ``\textit{I try to cold-call students equally from everyone, not only groups with cameras on or off... It does not matter. It is a killer mistake to assume those with cameras off less deserving or less intelligent. This is an unjustified prejudice we should avoid as teachers and remind ourselves not to fall prey to it.}'' (Associate Professor, Social Science)
\end{itemize}
\subsection{The Barrier List's Benefits for Novice Educators}
Finally, our summative study from study 3 shows that our barrier list can help novice educators in several different ways. Here, we summarize the three main benefits as follows.

\noindent\textbf{1. The barrier list can help novice educators not to forget about non-technological barriers.}
Through the survey responses, we notice that without-list educators tend to focus too much on solving the technological barriers, to the extent that they might sometimes forget about the other significant barriers students face.

According to the follow-up semi-structured interview of study 3 and the comparison of the survey responses of students and educators, we find that human-side and environmental barriers are often ignored or underappreciated by these educators. However, students attach much more importance to these two aspects, especially to human-side barriers. 
Interestingly, the later interviews reveal that most of these without-list educators were in fact aware of the non-technological barriers. However, they acknowledge three reasons for not thinking about such barriers when writing course plans: forgetfulness, misassumptions, and underestimation. 
\begin{itemize}
    \item [-] ``\textit{After all, faculties are humans too, and they might not know or remember every barrier students face. But I think it is in our nature to look at bigger and more tangible problems first. Technology is the biggest concern instructors and students deal with these days, and therefore it could naturally dominate the other kinds of barriers when we do course planning. [...] If I want to conclude, I would say that we need to collaborate more with senior colleagues and students to make our course plannings considerate and more authentic.}'' (Without-list group, Adjunct Instructor, Life Science)
    \item [-] ``\textit{I could name different reasons. The wrong assumption that we think old problems like human barriers or environmental problems have already been dealt with or resolve automatically by themselves is one of them. But in reality, old problems also grow and take new shapes and demand new kinds of supports. Although they might not look as exciting or as forceful as technological barriers, they can cripple the education if we just overlook them.}'' (Without-list group, Assistant Professor, Computer Science and Engineering)
\end{itemize}

\noindent\textbf{2. The barrier list can serve as a trustworthy reference for online course planning.} 
The analysis of study 3 also shows that without-list educators anticipate the student barriers mainly based on their own past experiences and intuition. Although this may help to deal with some very common teaching problems, many educators acknowledge that it is not a trustworthy reference for planning active online classes. Two major issues they mention in the follow-up interviews are the little experience they have with active online classes and the risk of missing or even incorrectly assuming some vital student barriers. Such misassumptions may worsen or even create more unnecessary barriers.
\begin{itemize}
    \item [-] ``\textit{We later might find we have made mistakes about barriers and that is a bad impression in early days of teaching. [...] A good reference can save your early days of teaching because it is the only time most of your student usually attend and they carefully evaluate everything to see if your class has worth to attend more sessions, turn on their camera, talk, participate and more ... if you leave a bad impression, they just take naps, go and show up at the end of the semester. For this reason, I think it is necessary to be informed about the possible barriers scientifically.}'' (Without-list group, Adjunct Instructor, Life Science)
\end{itemize}

\noindent\textbf{3. The barrier list can inspire educators to modify old active learning strategies and make them more inclusive.}
We notice that with-list educators tend to pose more creative solutions for addressing the student barriers. They are more active to suggest modifications to old active learning strategies, i.e., the ones we know from face-to-face classes. Many of these with-list educators describe the usage of the extracted list of barriers to be inspiring.
\begin{itemize}
    \item [-] ``\textit{One of the applications for this barrier list and maybe the similar ones could be inspiring faculty members to change their teaching styles. [...] The barrier list can also help [them] to feel more equipped to face new teaching challenges. [...] It is easy to go over the list in less than 10 minutes and get a quick idea about what is problematic for students.}'' (With-list group, Lecturer, Life Science)
    \item [-] ``\textit{Sometimes faculty members use heavy workload to make up for what is lost in online active classes. So they assign more homework. But what is ought to be done by faculty is not to just add more. They should rethink about now I am using a different way (of teaching)... not in a physical classroom, but in a synchronous online class. So how do I do what I used to do, differently, to achieve the same goal.}'' (With-list group, Professor, Electrical Engineering)
\end{itemize}
\section{Discussion}
In general, this study presents empirical evidence to the barriers students face in active online classes. We argue that this empirical evidence is important to the learning research community as many archived works in this domain are motivated by or deal with anecdotal evidence. This work contributes by bringing evidence from the reality of student barriers in active online classes, which has rarely been precisely documented.

We also help the community by correcting some misconceptions about AL. For example, while previous studies have mainly assumed AL to be automatically engendering positive feelings for class engagement, our study shows it to be a wrong assumption. According to our findings, apathy towards AL is one of the pervasive issues many students experience. This suggests that there might be a demand for designing the interface and interaction features for promoting positivity towards AL among students and for raising instructors' awareness of resistance in student participation.

Finally, the insights from this work can not only help the learning domain researchers and practitioners conduct a more informed inspection of the appropriateness of existing online education systems in AL settings, but also point out unmet needs as opportunities for future research and design. 

The rest of this section is organized into three parts. First, we provide some design suggestions for HCI and Learning at Scale researchers and practitioners. Next, our thematic analysis is elaborated, and finally, we introduce the main limitations of our work. 

\subsection{Design Suggestions}
\noindent$\bullet$ \textbf{Develop new interaction designs based on new input ways.} 
According to our barrier list, there are many barriers that make students unable or unwilling to turn on their cameras or microphones, and current pure text communication is too slow. Therefore, for the online education system, there is a demand for new interaction design based on new input ways, like screen sharing, Virtual Reality (VR), and Augmented Reality (AR). For example, interactive screen sharing is a very common way of transmitting information in online teaching \cite{Tee2006}, so students may use the shared screen to communicate with the teacher instantly and directly. A possible design could be that if a student has a question about one step of a formula derivation on a slide that is being shared, he or she can simply circle it with the mouse and mark it as ``question'' or ``unclear''. The action will be shared to the teacher's screen in real-time so that students can easily ask questions without having to type long sentences and weird symbols in the chat box or explicitly pop up and interrupt the teacher, which are also barriers that decrease the students' willingness to engage, especially in large-scale classes where some students would be reluctant to directly ask questions due to shyness and low self-efficacy. Using such new types of input in the interaction process utilizes the advantage of online education and improves communication efficiency between students and educators.

\noindent$\bullet$ \textbf{Apply virtualization techniques to transmit more information in communication under the premise of adequate privacy protection.} 
There is always a trade-off between privacy and online communication efficiency. Although it is common sense that turning on the camera can significantly increase communication efficiency, a student often refuses to do so for privacy reasons, such as unwillingness to reveal his or her messy room or casual dress or face without makeup, and so on \cite{Hasan2019}. Some techniques such as virtual background have been used to address these barriers but it only solves part of the problem. What we further recommend is to use virtual faces such as Memoji or Animoji to better protect the privacy of the user without vital information loss. This is because virtual faces can replace the original face with no loss of facial expression such as smile, gaze, frown, nod, and shake which contain emotional information \cite{Bacos2020, Muralidhar2018}. Such virtual interactive design can alleviate people's privacy concerns so that encourage more people to turn the camera on, which will greatly increase communication efficiency. Hence, more research in this direction is demanded.

\noindent$\bullet$ \textbf{Instant communication between students might help with students' learning.}
Although excessive whispering in the classroom is considered bad behavior with disrespect, it is believed that timely and appropriate private communication is necessary and beneficial for learning especially in online classes \cite{yardi2008whispers}. In our interviews, many students mentioned that the lack of instant communication makes it difficult for them to keep up with the class once they miss any important information. Meanwhile, some students had to use external social media apps to communicate with their classmates in class, yet it was easy to be distracted by these apps. Hence, we believe that a function of allowing whispering among students should also exist in online education systems. However, in current online education systems, functionalities for students to communicate with each other privately is still totally text-based and usually disabled by the instructor. Therefore, we encourage HCI and Learning at Scale researchers to explore more efficient designs for instant communication in the online education system, enabling students to communicate in a timely and efficient manner.

\subsection{Thematic Analysis}
In this research, we use an integrated deductive and inductive coding approach to analyze the extracted qualitative data from the social media content and all of the transcripts of the semi-structured interviews in studies 1 and 2 \cite{kyngas2019application}. The deductive part is because we have clearly defined the scope of this work as exploring student barriers to active online learning right from the beginning, and there exist some frameworks of offline active learning barriers that could serve as initial seeds of our investigation (see \cite{michael2007faculty}). Meanwhile, we realize that there might be emerging issues in active online learning, which might be different from the offline counterpart as indicated in the POD network discussions of faculty members, which motivated this work as presented in the introduction. We thus introduce the inductive component of analysis to keep the room open for any type of barriers that emerge directly out of our data sources. 

Towards this end, this research follows a standard coding procedure \cite{willson2019analysing}. Different members first read all of the extracted content and transcripts multiple times and then extract the initial themes and codes independently. The findings are merged and refined iteratively through weekly group meetings and discussions for over two months. We use affinity clustering diagrams shared with Google sheets to save, manage, and organize our findings. We achieve \textit{Fleiss’ Kappa} inter-rater reliability measure of 0.89 on our final list of student barriers, which shows an almost perfect consensus among all of the team members \cite{10.1145/2998181.2998342}.

\subsection{Limitations}
We should acknowledge some limitations in relation to our results. Since we used convenience and snowball sampling to recruit participants, and although we tried to ensure the diversity of our participants’ backgrounds, there might still be some bias. For example, more than half of the interviewees were from engineering majors, and the majority of the educators were male. We also found it hard to find educators who have experience in active online classes, because most universities have only been conducting online teaching for about two semesters. Therefore, we plan to collaborate with universities and researchers around the world for further in-depth research. Moreover, due to the pandemic, all interviews have been conducted online and we thus had less contact with our participants; so some non-verbal cues may have been missed. Therefore, we think it would be advantageous to interview more participants face-to-face in future follow-up studies.

\section{Conclusion}
In this paper, we adopt a qualitative research approach to characterize student barriers to active learning in synchronous online environments. Opinions of students and expert educators were collected from social media content and semi-structured interviews, and then thematically analyzed to create a nuanced list of student barriers within the themes of human-side, technological, and environmental barriers.
Next, we conducted surveys with students and expert educators who had prior experience with active learning courses to explore the importance and exclusiveness of each barrier. Finally, the results of a summative study with novice educators showed the usefulness and effectiveness of our barrier list. This work aims to help novice educators comprehensively consider student barriers and prepare more student-centered and inclusive active online classes. Furthermore, this is the first study of student barriers to draw HCI and Learning at Scale researchers’ attention to active learning in synchronous online classes. However, our research was mainly based on local universities and therefore, future work should investigate students and educators from more diverse backgrounds and on a larger scale.

\section{Acknowledgment}
Here, we want to thank all of the research participants who kindly took part in our work and made it possible. This research has been supported in part by project 16214817 from the Research Grants Council of Hong Kong, and the 5GEAR and FIT projects from Academy of Finland. It is also partially supported by funding from the Theme-based Research Scheme of the Hong Kong Research Grants Council, grant no. T44-707/16-N.
\balance{}

\bibliographystyle{SIGCHI-Reference-Format}
\bibliography{sample}


\begin{thebibliography}{00}


\ifx \showCODEN    \undefined \def \showCODEN     #1{\unskip}     \fi
\ifx \showDOI      \undefined \def \showDOI       #1{{\tt DOI:}\penalty0{#1}\ }
  \fi
\ifx \showISBNx    \undefined \def \showISBNx     #1{\unskip}     \fi
\ifx \showISBNxiii \undefined \def \showISBNxiii  #1{\unskip}     \fi
\ifx \showISSN     \undefined \def \showISSN      #1{\unskip}     \fi
\ifx \showLCCN     \undefined \def \showLCCN      #1{\unskip}     \fi
\ifx \shownote     \undefined \def \shownote      #1{#1}          \fi
\ifx \showarticletitle \undefined \def \showarticletitle #1{#1}   \fi
\ifx \showURL      \undefined \def \showURL       #1{#1}          \fi

\bibitem{abdelsattar2019active}
{Amal AbdelSattar} {and} {Wafa Labib}. 2019.
\newblock \showarticletitle{Active Learning in Engineering Education: Teaching
  Strategies and Methods of Overcoming Challenges}. In {\em Proceedings of the
  2019 8th International Conference on Educational and Information Technology}.
  255--261.
\newblock


\bibitem{aksit2016active}
{Fisun Aksit}, {Hannele Niemi}, {and} {Anne Nevgi}. 2016.
\newblock \showarticletitle{Why is active learning so difficult to implement:
  The Turkish case}.
\newblock {\em Australian Journal of Teacher Education\/} {41}, 4 (2016), 6.
\newblock


\bibitem{10.1145/3313831.3376776}
{Mehdi Alaimi}, {Edith Law}, {Kevin~Daniel Pantasdo}, {Pierre-Yves Oudeyer},
  {and} {H\'{e}l\`{e}ne Sauzeon}. 2020.
\newblock \showarticletitle{Pedagogical Agents for Fostering Question-Asking
  Skills in Children}. In {\em Proceedings of the 2020 CHI Conference on Human
  Factors in Computing Systems} {\em (CHI ’20)}. Association for Computing
  Machinery, New York, NY, USA, 1–13.
\newblock
\showISBNx{9781450367080}
\showDOI{%
\url{http://dx.doi.org/10.1145/3313831.3376776}}


\bibitem{10.1145/3290605.3300534}
{Sinem Aslan}, {Nese Alyuz}, {Cagri Tanriover}, {Sinem~E. Mete}, {Eda Okur},
  {Sidney~K. D’Mello}, {and} {Asli Arslan~Esme}. 2019.
\newblock \showarticletitle{Investigating the Impact of a Real-Time, Multimodal
  Student Engagement Analytics Technology in Authentic Classrooms}. In {\em
  Proceedings of the 2019 CHI Conference on Human Factors in Computing Systems}
  {\em (CHI ’19)}. Association for Computing Machinery, New York, NY, USA,
  1–12.
\newblock
\showISBNx{9781450359702}
\showDOI{%
\url{http://dx.doi.org/10.1145/3290605.3300534}}


\bibitem{Bacos2020}
{Catherine~A. Bacos}. 2020.
\newblock \showarticletitle{Enhancing Social Attention Using Eye-Movement
  Modeling and Simulated Dyadic Social Interactions}. In {\em Extended
  Abstracts of the 2020 CHI Conference on Human Factors in Computing Systems}
  {\em (CHI EA '20)}. Association for Computing Machinery, New York, NY, USA,
  1–6.
\newblock
\showISBNx{9781450368193}
\showDOI{%
\url{http://dx.doi.org/10.1145/3334480.3381448}}


\bibitem{ballantyne2007documenting}
{Julie Ballantyne}. 2007.
\newblock \showarticletitle{Documenting praxis shock in early-career Australian
  music teachers: The impact of pre-service teacher education}.
\newblock {\em International Journal of Music Education\/} {25}, 3 (2007),
  181--191.
\newblock


\bibitem{10.1145/3313831.3376518}
{Jonathan Bassen}, {Bharathan Balaji}, {Michael Schaarschmidt}, {Candace
  Thille}, {Jay Painter}, {Dawn Zimmaro}, {Alex Games}, {Ethan Fast}, {and}
  {John~C. Mitchell}. 2020.
\newblock \showarticletitle{Reinforcement Learning for the Adaptive Scheduling
  of Educational Activities}. In {\em Proceedings of the 2020 CHI Conference on
  Human Factors in Computing Systems} {\em (CHI ’20)}. Association for
  Computing Machinery, New York, NY, USA, 1–12.
\newblock
\showISBNx{9781450367080}
\showDOI{%
\url{http://dx.doi.org/10.1145/3313831.3376518}}


\bibitem{bonwell1991active}
{Charles~C Bonwell} {and} {James~A Eison}. 1991.
\newblock {\em Active Learning: Creating Excitement in the Classroom. 1991
  ASHE-ERIC Higher Education Reports.}
\newblock ERIC.
\newblock


\bibitem{bonwell1996active}
{Charles~C Bonwell} {and} {Tracey~E Sutherland}. 1996.
\newblock \showarticletitle{The active learning continuum: Choosing activities
  to engage students in the classroom}.
\newblock {\em New directions for teaching and learning\/} {1996}, 67 (1996),
  3--16.
\newblock


\bibitem{bowen2017understanding}
{Ryan~S Bowen}. 2017.
\newblock \showarticletitle{Understanding by Design. Vanderbilt University
  Center for Teaching}.
\newblock {\em Retrieved April\/}  {25} (2017), 2019.
\newblock


\bibitem{budhai2017best}
{Stephanie~Smith Budhai} {and} {others}. 2017.
\newblock {\em Best practices in engaging online learners through active and
  experiential learning strategies}.
\newblock Taylor \& Francis.
\newblock


\bibitem{Cplanner2017}
{Understanding by Design}. 2017.
\newblock Course Planner Template.
\newblock   (August 2017).
\newblock
\showURL{%
Retrieved August 12, 2020 from
  \url{https://cft.vanderbilt.edu/guides-sub-pages/understanding-by-design/}}


\bibitem{morethanzoom}
{Kevin Carey}. 2020.
\newblock Everybody Ready for the Big Migration to Online College? Actually,
  No.
\newblock   (July 2020).
\newblock
\showURL{%
Retrieved July 11, 2020 from
  \url{https://www.nytimes.com/2020/03/13/upshot/coronavirus-online-college-classes-unprepared.html}}


\bibitem{cattaneo2017telling}
{Kelsey~Hood Cattaneo}. 2017.
\newblock \showarticletitle{Telling active learning pedagogies apart: From
  theory to practice}.
\newblock {\em Journal of New Approaches in Educational Research (NAER
  Journal)\/} {6}, 2 (2017), 144--152.
\newblock


\bibitem{clark2015comparing}
{Cynthia Clark}, {Neal Strudler}, {and} {Karen Grove}. 2015.
\newblock \showarticletitle{Comparing asynchronous and synchronous video vs.
  text based discussions in an online teacher education course.}
\newblock {\em Online Learning\/} {19}, 3 (2015), 48--69.
\newblock


\bibitem{cohen2019think}
{Matthew Cohen}, {Steven~G Buzinski}, {Emma Armstrong-Carter}, {Jenna Clark},
  {Benjamin Buck}, {and} {Lillian Reuman}. 2019.
\newblock \showarticletitle{Think, pair, freeze: The association between social
  anxiety and student discomfort in the active learning environment.}
\newblock {\em Scholarship of Teaching and Learning in Psychology\/} (2019).
\newblock


\bibitem{10.1145/3290605.3300368}
{Amy Cook}, {Jessica Hammer}, {Salma Elsayed-Ali}, {and} {Steven Dow}. 2019.
\newblock \showarticletitle{How Guiding Questions Facilitate Feedback Exchange
  in Project-Based Learning}. In {\em Proceedings of the 2019 CHI Conference on
  Human Factors in Computing Systems} {\em (CHI '19)}. Association for
  Computing Machinery, New York, NY, USA, 1–12.
\newblock
\showISBNx{9781450359702}
\showDOI{%
\url{http://dx.doi.org/10.1145/3290605.3300368}}


\bibitem{cooper2020investigating}
{Haley Cooper}. 2020.
\newblock \showarticletitle{Investigating Active Learning in Inclusion and
  Resource Language Arts Classrooms}.
\newblock  (2020).
\newblock


\bibitem{csikszentmihalyi2020finding}
{Mihaly Csikszentmihalyi}. 2020.
\newblock {\em Finding flow: The psychology of engagement with everyday life}.
\newblock Hachette UK.
\newblock


\bibitem{de2019let}
{Adriano~Luiz de Souza~Lima} {and} {Fabiane Barreto~Vavassori Benitti}. 2019.
\newblock \showarticletitle{Let’s Talk About Tools and Approaches for
  Teaching HCI}. In {\em International Conference on Human-Computer
  Interaction}. Springer, 155--170.
\newblock


\bibitem{10.1145/2207676.2208358}
{Tao Dong}, {Mira Dontcheva}, {Diana Joseph}, {Karrie Karahalios}, {Mark
  Newman}, {and} {Mark Ackerman}. 2012.
\newblock \showarticletitle{Discovery-Based Games for Learning Software}. In
  {\em Proceedings of the SIGCHI Conference on Human Factors in Computing
  Systems} {\em (CHI '12)}. Association for Computing Machinery, New York, NY,
  USA, 2083–2086.
\newblock
\showISBNx{9781450310154}
\showDOI{%
\url{http://dx.doi.org/10.1145/2207676.2208358}}


\bibitem{edwards2017like}
{Susan Edwards}. 2017.
\newblock \showarticletitle{Like a chameleon: A beginning teacher’s journey
  to implement active learning}.
\newblock {\em RMLE Online\/} {40}, 4 (2017), 1--11.
\newblock


\bibitem{10.1145/2998181.2998342}
{Sarah Evans}, {Katie Davis}, {Abigail Evans}, {Julie~Ann Campbell}, {David~P.
  Randall}, {Kodlee Yin}, {and} {Cecilia Aragon}. 2017.
\newblock \showarticletitle{More Than Peer Production: Fanfiction Communities
  as Sites of Distributed Mentoring}. In {\em Proceedings of the 2017 ACM
  Conference on Computer Supported Cooperative Work and Social Computing} {\em
  (CSCW '17)}. Association for Computing Machinery, New York, NY, USA,
  259–272.
\newblock
\showISBNx{9781450343350}
\showDOI{%
\url{http://dx.doi.org/10.1145/2998181.2998342}}


\bibitem{faulconer2018comparison}
{Emily~K Faulconer}, {J Griffith}, {Beverly Wood}, {S Acharyya}, {and} {D
  Roberts}. 2018.
\newblock \showarticletitle{A comparison of online, video synchronous, and
  traditional learning modes for an introductory undergraduate physics course}.
\newblock {\em Journal of Science Education and Technology\/} {27}, 5 (2018),
  404--411.
\newblock


\bibitem{freeman2014active}
{Scott Freeman}, {Sarah~L Eddy}, {Miles McDonough}, {Michelle~K Smith},
  {Nnadozie Okoroafor}, {Hannah Jordt}, {and} {Mary~Pat Wenderoth}. 2014.
\newblock \showarticletitle{Active learning increases student performance in
  science, engineering, and mathematics}.
\newblock {\em Proceedings of the National Academy of Sciences\/} {111}, 23
  (2014), 8410--8415.
\newblock


\bibitem{garrison1999critical}
{D~Randy Garrison}, {Terry Anderson}, {and} {Walter Archer}. 1999.
\newblock \showarticletitle{Critical inquiry in a text-based environment:
  Computer conferencing in higher education}.
\newblock {\em The internet and higher education\/} {2}, 2-3 (1999), 87--105.
\newblock


\bibitem{ghergulescu2019interactive}
{Ioana Ghergulescu}, {Arghir-Nicolae Moldovan}, {Cristina~Hava Muntean}, {and}
  {Gabriel-Miro Muntean}. 2019.
\newblock \showarticletitle{Interactive Personalised STEM Virtual Lab Based on
  Self-Directed Learning and Self-Efficacy}. In {\em Adjunct Publication of the
  27th Conference on User Modeling, Adaptation and Personalization}. 355--358.
\newblock


\bibitem{goldberg2012active}
{Jay~R Goldberg}. 2012.
\newblock \showarticletitle{Active Learning in Capstone Design Courses [Senior
  Design]}.
\newblock {\em IEEE pulse\/} {3}, 3 (2012), 54--57.
\newblock


\bibitem{guo2017mining}
{Yue Guo}, {Stuart~J Barnes}, {and} {Qiong Jia}. 2017.
\newblock \showarticletitle{Mining meaning from online ratings and reviews:
  Tourist satisfaction analysis using latent dirichlet allocation}.
\newblock {\em Tourism Management\/}  {59} (2017), 467--483.
\newblock


\bibitem{harmon2017professional}
{Melissa~Cameron Harmon}. 2017.
\newblock {\em Professional Development as a Catalyst for Change in the
  Community College Science Classroom: How Active Learning Pedagogy Impacts
  Teaching Practices as Well as Faculty and Student Perceptions of Learning}.
\newblock Ph.D. Dissertation. Wingate University.
\newblock


\bibitem{Hasan2019}
{Rakibul Hasan}, {Yifang Li}, {Eman Hassan}, {Kelly Caine}, {David~J.
  Crandall}, {Roberto Hoyle}, {and} {Apu Kapadia}. 2019.
\newblock \showarticletitle{Can Privacy Be Satisfying? On Improving Viewer
  Satisfaction for Privacy-Enhanced Photos Using Aesthetic Transforms}. In {\em
  Proceedings of the 2019 CHI Conference on Human Factors in Computing Systems}
  {\em (CHI '19)}. Association for Computing Machinery, New York, NY, USA,
  1–13.
\newblock
\showISBNx{9781450359702}
\showDOI{%
\url{http://dx.doi.org/10.1145/3290605.3300597}}


\bibitem{henriksen2020folk}
{Danah Henriksen}, {Edwin Creely}, {and} {Michael Henderson}. 2020.
\newblock \showarticletitle{Folk pedagogies for teacher transitions: Approaches
  to synchronous online learning in the wake of COVID-19}.
\newblock {\em Journal of Technology and Teacher Education\/} {28}, 2 (2020),
  201--209.
\newblock


\bibitem{hollis2016mind}
{R~Benjamin Hollis} {and} {Christopher~A Was}. 2016.
\newblock \showarticletitle{Mind wandering, control failures, and social media
  distractions in online learning}.
\newblock {\em Learning and Instruction\/}  {42} (2016), 104--112.
\newblock


\bibitem{huang2020handbook}
{RH Huang}, {DJ Liu}, {A Tlili}, {JF Yang}, {HH Wang}, {and} {others}. 2020.
\newblock \showarticletitle{Handbook on facilitating flexible learning during
  educational disruption: The Chinese experience in maintaining undisrupted
  learning in COVID-19 Outbreak}.
\newblock {\em Beijing: Smart Learning Institute of Beijing Normal
  University\/} (2020).
\newblock


\bibitem{10.1145/2858036.2858184}
{Clarissa Ishak}, {Carman Neustaedter}, {Dan Hawkins}, {Jason Procyk}, {and}
  {Michael Massimi}. 2016.
\newblock \showarticletitle{Human Proxies for Remote University Classroom
  Attendance}. In {\em Proceedings of the 2016 CHI Conference on Human Factors
  in Computing Systems} {\em (CHI ’16)}. Association for Computing Machinery,
  New York, NY, USA, 931–943.
\newblock
\showISBNx{9781450333627}
\showDOI{%
\url{http://dx.doi.org/10.1145/2858036.2858184}}


\bibitem{10.1145/3313831.3376418}
{Emily Jensen}, {Meghan Dale}, {Patrick~J. Donnelly}, {Cathlyn Stone}, {Sean
  Kelly}, {Amanda Godley}, {and} {Sidney~K. D’Mello}. 2020.
\newblock \showarticletitle{Toward Automated Feedback on Teacher Discourse to
  Enhance Teacher Learning}. In {\em Proceedings of the 2020 CHI Conference on
  Human Factors in Computing Systems} {\em (CHI ’20)}. Association for
  Computing Machinery, New York, NY, USA, 1–13.
\newblock
\showISBNx{9781450367080}
\showDOI{%
\url{http://dx.doi.org/10.1145/3313831.3376418}}


\bibitem{kock2005media}
{Ned Kock}. 2005.
\newblock \showarticletitle{Media richness or media naturalness? The evolution
  of our biological communication apparatus and its influence on our behavior
  toward e-communication tools}.
\newblock {\em IEEE transactions on professional communication\/} {48}, 2
  (2005), 117--130.
\newblock


\bibitem{kock2011media}
{Ned Kock}. 2011.
\newblock \showarticletitle{Media naturalness theory: human evolution and
  behaviour towards}.
\newblock {\em Applied evolutionary psychology\/}  {381} (2011).
\newblock


\bibitem{10.1145/3330430.3333653}
{Denise~G. Kutnick} {and} {David~A. Joyner}. 2019.
\newblock \showarticletitle{Synchronous at Scale: Investigation and
  Implementation of a Semi-Synchronous Online Lecture Platform}. In {\em
  Proceedings of the Sixth (2019) ACM Conference on Learning @ Scale} {\em (L@S
  '19)}. Association for Computing Machinery, New York, NY, USA, Article 40, 4
  pages.
\newblock
\showISBNx{9781450368049}
\showDOI{%
\url{http://dx.doi.org/10.1145/3330430.3333653}}


\bibitem{kyngas2019application}
{Helvi Kyng{\"a}s}, {Kristina Mikkonen}, {and} {Maria K{\"a}{\"a}ri{\"a}inen}.
  2019.
\newblock {\em The application of content analysis in nursing science
  research}.
\newblock Springer.
\newblock


\bibitem{morethan80}
{Doug Lederman}. 2019.
\newblock Provosts Count More on Online Programs.
\newblock   (june 2019).
\newblock
\showURL{%
Retrieved June 22, 2020 from
  \url{https://www.insidehighered.com/digital-learning/article/2019/01/23/provosts-aim-lean-more-heavily-online-programs}}


\bibitem{10.1145/2702123.2702349}
{Yi-Chieh Lee}, {Wen-Chieh Lin}, {Fu-Yin Cherng}, {Hao-Chuan Wang}, {Ching-Ying
  Sung}, {and} {Jung-Tai King}. 2015.
\newblock \showarticletitle{Using Time-Anchored Peer Comments to Enhance Social
  Interaction in Online Educational Videos}. In {\em Proceedings of the 33rd
  Annual ACM Conference on Human Factors in Computing Systems} {\em (CHI
  ’15)}. Association for Computing Machinery, New York, NY, USA, 689–698.
\newblock
\showISBNx{9781450331456}
\showDOI{%
\url{http://dx.doi.org/10.1145/2702123.2702349}}


\bibitem{li2010inquiry}
{Qing Li}, {Lynn Moorman}, {and} {Patti Dyjur}. 2010.
\newblock \showarticletitle{Inquiry-based learning and e-mentoring via
  videoconference: a study of mathematics and science learning of Canadian
  rural students}.
\newblock {\em Educational Technology Research and Development\/} {58}, 6
  (2010), 729--753.
\newblock


\bibitem{lindvall2019coherence}
{Jannika Lindvall} {and} {Andreas Ryve}. 2019.
\newblock \showarticletitle{Coherence and the positioning of teachers in
  professional development programs. A systematic review}.
\newblock {\em Educational Research Review\/}  {27} (2019), 140--154.
\newblock


\bibitem{loughran2016teaching}
{John Loughran}. 2016.
\newblock \showarticletitle{Teaching and teacher education: The need to go
  beyond rhetoric}.
\newblock In {\em Teacher education}. Springer, 253--264.
\newblock


\bibitem{martin2019award}
{Florence Martin}, {Albert Ritzhaupt}, {Swapna Kumar}, {and} {Kiran Budhrani}.
  2019.
\newblock \showarticletitle{Award-winning faculty online teaching practices:
  Course design, assessment and evaluation, and facilitation}.
\newblock {\em The Internet and Higher Education\/}  {42} (2019), 34--43.
\newblock


\bibitem{michael2007faculty}
{Joel Michael}. 2007.
\newblock \showarticletitle{Faculty perceptions about barriers to active
  learning}.
\newblock {\em College teaching\/} {55}, 2 (2007), 42--47.
\newblock


\bibitem{mintzes2020constructivism}
{Joel~J Mintzes}. 2020.
\newblock \showarticletitle{From constructivism to active learning in college
  science}.
\newblock In {\em Active Learning in College Science}. Springer, 3--12.
\newblock


\bibitem{mintzes2020active}
{Joel~J Mintzes} {and} {Emily~M Walter}. 2020.
\newblock Active learning in college science: The case for evidence-based
  practice.
\newblock   (2020).
\newblock


\bibitem{mogavi2019hrcr}
{Reza~Hadi Mogavi}, {Sujit Gujar}, {Xiaojuan Ma}, {and} {Pan Hui}. 2019.
\newblock \showarticletitle{HRCR: Hidden Markov-based Reinforcement to Reduce
  Churn in Question Answering Forums}. In {\em Pacific Rim International
  Conference on Artificial Intelligence}. Springer, 364--376.
\newblock


\bibitem{mogavipoolwebsites}
{Reza~Hadi Mogavi}, {Xiaojuan Ma}, {and} {Pan Hui}. 2021.
\newblock \showarticletitle{Characterizing Student Engagement Moods for Dropout
  Prediction in Question Pool Websites}. In {\em In Proceedings of the 24th ACM
  Conference on Computer-Supported Cooperative Work and Social Computing} {\em
  (CSCW '21)}. Association for Computing Machinery, New York, NY, USA, 22.
\newblock
\showDOI{%
\url{http://dx.doi.org/10.1145/3449086}}


\bibitem{Muralidhar2018}
{Skanda Muralidhar}, {R\'{e}my Siegfried}, {Jean-Marc Odobez}, {and} {Daniel
  Gatica-Perez}. 2018.
\newblock \showarticletitle{Facing Employers and Customers: What Do Gaze and
  Expressions Tell About Soft Skills?}. In {\em Proceedings of the 17th
  International Conference on Mobile and Ubiquitous Multimedia} {\em (MUM
  2018)}. Association for Computing Machinery, New York, NY, USA, 121–126.
\newblock
\showISBNx{9781450365949}
\showDOI{%
\url{http://dx.doi.org/10.1145/3282894.3282925}}


\bibitem{naeem2020analyzing}
{Ayesha Naeem~Syeda}, {Rutwa Engineer}, {and} {Bogdan Simion}. 2020.
\newblock \showarticletitle{Analyzing the Effects of Active Learning Classrooms
  in CS2}. In {\em Proceedings of the 51st ACM Technical Symposium on Computer
  Science Education}. 93--99.
\newblock


\bibitem{PODpeople1}
{POD Network}. 2020.
\newblock POD Network Open Discussion Group.
\newblock   (July 2020).
\newblock
\showURL{%
Retrieved July 20, 2020 from
  \url{https://groups.google.com/a/podnetwork.org/g/discussion}}


\bibitem{nicol2018comparison}
{Adelheid~AM Nicol}, {Soo~M Owens}, {St{\'e}phanie~SCL Le~Coze}, {Allister
  MacIntyre}, {and} {Christina Eastwood}. 2018.
\newblock \showarticletitle{Comparison of high-technology active learning and
  low-technology active learning classrooms}.
\newblock {\em Active Learning in Higher Education\/} {19}, 3 (2018), 253--265.
\newblock


\bibitem{10.1145/3313831.3376149}
{Alannah Oleson}, {Meron Solomon}, {and} {Amy~J. Ko}. 2020.
\newblock \showarticletitle{Computing Students’ Learning Difficulties in HCI
  Education}. In {\em Proceedings of the 2020 CHI Conference on Human Factors
  in Computing Systems} {\em (CHI ’20)}. Association for Computing Machinery,
  New York, NY, USA, 1–14.
\newblock
\showISBNx{9781450367080}
\showDOI{%
\url{http://dx.doi.org/10.1145/3313831.3376149}}


\bibitem{ouh2019applying}
{Eng~Lieh Ouh} {and} {Yunghans Irawan}. 2019.
\newblock \showarticletitle{Applying case-based learning for a postgraduate
  software architecture course}. In {\em Proceedings of the 2019 ACM Conference
  on Innovation and Technology in Computer Science Education}. 457--463.
\newblock


\bibitem{pan2020learning}
{Zilong Pan}, {Chenglu Li}, {and} {Min Liu}. 2020.
\newblock \showarticletitle{Learning Analytics Dashboard for Problem-based
  Learning}. In {\em Proceedings of the Seventh ACM Conference on Learning@
  Scale}. 393--396.
\newblock


\bibitem{UbsuggestEngine}
{Neil Patel}. 2020.
\newblock Ubersuggest's Free Keyword Tool, Generate More Suggestions.
\newblock   (April 2020).
\newblock
\showURL{%
Retrieved April 15, 2020 from \url{https://neilpatel.com/ubersuggest/}}


\bibitem{phelps2019successful}
{Amber Phelps} {and} {Dimitrios Vlachopoulos}. 2019.
\newblock \showarticletitle{Successful transition to synchronous learning
  environments in distance education: A research on entry-level synchronous
  facilitator competencies}.
\newblock {\em Education and Information Technologies\/} (2019), 1--17.
\newblock


\bibitem{piskurich2009rapid}
{George~M Piskurich}. 2009.
\newblock {\em Rapid training development: Developing training courses fast and
  right}.
\newblock John Wiley \& Sons.
\newblock


\bibitem{polleck2020putting}
{Jody Polleck} {and} {Jordan Yarwood}. 2020.
\newblock \showarticletitle{Putting students at the center: Empowering urban
  alternative education teachers through culturally relevant and sustaining
  unit planning}.
\newblock {\em Preventing School Failure: Alternative Education for Children
  and Youth\/} {64}, 3 (2020), 191--200.
\newblock


\bibitem{potluri2018codetalk}
{Venkatesh Potluri}, {Priyan Vaithilingam}, {Suresh Iyengar}, {Y Vidya},
  {Manohar Swaminathan}, {and} {Gopal Srinivasa}. 2018.
\newblock \showarticletitle{Codetalk: Improving programming environment
  accessibility for visually impaired developers}. In {\em Proceedings of the
  2018 CHI Conference on Human Factors in Computing Systems}. 1--11.
\newblock


\bibitem{preece2015interaction}
{Jennifer Preece}, {Helen Sharp}, {and} {Yvonne Rogers}. 2015.
\newblock {\em Interaction design: beyond human-computer interaction}.
\newblock John Wiley \& Sons.
\newblock


\bibitem{prince2004does}
{Michael Prince}. 2004.
\newblock \showarticletitle{Does active learning work? A review of the
  research}.
\newblock {\em Journal of engineering education\/} {93}, 3 (2004), 223--231.
\newblock


\bibitem{10.1145/3290605.3300774}
{Iulian Radu} {and} {Bertrand Schneider}. 2019.
\newblock \showarticletitle{What Can We Learn from Augmented Reality (AR)?
  Benefits and Drawbacks of AR for Inquiry-Based Learning of Physics}. In {\em
  Proceedings of the 2019 CHI Conference on Human Factors in Computing Systems}
  {\em (CHI '19)}. Association for Computing Machinery, New York, NY, USA,
  1–12.
\newblock
\showISBNx{9781450359702}
\showDOI{%
\url{http://dx.doi.org/10.1145/3290605.3300774}}


\bibitem{10.1145/3313831.3376650}
{Dan Richardson} {and} {Ahmed Kharrufa}. 2020.
\newblock \showarticletitle{We Are the Greatest Showmen: Configuring a
  Framework for Project-Based Mobile Learning}. In {\em Proceedings of the 2020
  CHI Conference on Human Factors in Computing Systems} {\em (CHI '20)}.
  Association for Computing Machinery, New York, NY, USA, 1–12.
\newblock
\showISBNx{9781450367080}
\showDOI{%
\url{http://dx.doi.org/10.1145/3313831.3376650}}


\bibitem{schwartz2019bite}
{Ann~C Schwartz}, {Robert~O Cotes}, {Jungjin Kim}, {Martha~C Ward}, {and}
  {Kimberly~D Manning}. 2019.
\newblock \showarticletitle{Bite-Sized Teaching: Engaging the Modern Learner in
  Psychiatry}.
\newblock {\em Academic Psychiatry\/} {43}, 3 (2019), 315--318.
\newblock


\bibitem{10.1145/3027063.3053148}
{Jinsil~Hwaryoung Seo}, {Brian~Michael Smith}, {Margaret~E. Cook}, {Erica~R.
  Malone}, {Michelle Pine}, {Steven Leal}, {Zhikun Bai}, {and} {Jinkyo Suh}.
  2017.
\newblock \showarticletitle{Anatomy Builder VR: Embodied VR Anatomy Learning
  Program to Promote Constructionist Learning}. In {\em Proceedings of the 2017
  CHI Conference Extended Abstracts on Human Factors in Computing Systems} {\em
  (CHI EA ’17)}. Association for Computing Machinery, New York, NY, USA,
  2070–2075.
\newblock
\showISBNx{9781450346566}
\showDOI{%
\url{http://dx.doi.org/10.1145/3027063.3053148}}


\bibitem{settles2009active}
{Burr Settles}. 2009.
\newblock \showarticletitle{Active learning literature survey}.
\newblock  (2009).
\newblock


\bibitem{sharifrazi2019students}
{Farnaz Sharifrazi} {and} {Suki Stone}. 2019.
\newblock \showarticletitle{Students Perception of Learning Online: Professor's
  Presence in Synchronous Versus Asynchronous Modality}. In {\em Proceedings of
  the 2019 5th International Conference on Computer and Technology
  Applications}. 180--183.
\newblock


\bibitem{silverthorn2020active}
{Dee~Unglaub Silverthorn}. 2020.
\newblock \showarticletitle{When Active Learning Fails… and What to Do About
  It}.
\newblock In {\em Active Learning in College Science}. Springer, 985--1001.
\newblock


\bibitem{stuart1978medical}
{John Stuart} {and} {RJ~Desmond Rutherford}. 1978.
\newblock \showarticletitle{Medical student concentration during lectures}.
\newblock {\em The lancet\/} {312}, 8088 (1978), 514--516.
\newblock


\bibitem{10.1145/3290605.3300662}
{Na Sun}, {Xiying Wang}, {and} {Mary~Beth Rosson}. 2019.
\newblock \showarticletitle{How Do Distance Learners Connect?}. In {\em
  Proceedings of the 2019 CHI Conference on Human Factors in Computing Systems}
  {\em (CHI ’19)}. Association for Computing Machinery, New York, NY, USA,
  1–12.
\newblock
\showISBNx{9781450359702}
\showDOI{%
\url{http://dx.doi.org/10.1145/3290605.3300662}}


\bibitem{tantasawat2019attitudes}
{Piyada~Alisha Tantasawat}, {Sutthinee Srisawat}, {Narudol Damsugree},
  {Amornthep Thepwichit}, {and} {Panlada Tittabutr}. 2019.
\newblock \showarticletitle{Attitudes Toward Using E-Courseware in A Flipped
  Classroom Teaching And Learning Approach of Suranaree University of
  Technology Students in The Application Of Biotechnology In Crop Production
  Course}. In {\em Proceedings of the 2019 3rd International Conference on
  Education and Multimedia Technology}. 206--214.
\newblock


\bibitem{Tee2006}
{Kimberly Tee}, {Saul Greenberg}, {and} {Carl Gutwin}. 2006.
\newblock \showarticletitle{Providing Artifact Awareness to a Distributed Group
  through Screen Sharing}. In {\em Proceedings of the 2006 20th Anniversary
  Conference on Computer Supported Cooperative Work} {\em (CSCW '06)}.
  Association for Computing Machinery, New York, NY, USA, 99–108.
\newblock
\showISBNx{1595932496}
\showDOI{%
\url{http://dx.doi.org/10.1145/1180875.1180891}}


\bibitem{tharayil2018strategies}
{Sneha Tharayil}, {Maura Borrego}, {Michael Prince}, {Kevin~A Nguyen}, {Prateek
  Shekhar}, {Cynthia~J Finelli}, {and} {Cynthia Waters}. 2018.
\newblock \showarticletitle{Strategies to mitigate student resistance to active
  learning}.
\newblock {\em International Journal of STEM Education\/} {5}, 1 (2018), 7.
\newblock


\bibitem{thomas1972variation}
{EJ Thomas}. 1972.
\newblock \showarticletitle{The variation of memory with time for information
  appearing during a lecture}.
\newblock {\em Studies in adult education\/} {4}, 1 (1972), 57--62.
\newblock


\bibitem{10.1145/3270316.3270611}
{Ruth Torres~Castillo}. 2018.
\newblock \showarticletitle{Computer Games As Learning Tools: Teachers
  Attitudes \& Behaviors}. In {\em Proceedings of the 2018 Annual Symposium on
  Computer-Human Interaction in Play Companion Extended Abstracts} {\em (CHI
  PLAY ’18 Extended Abstracts)}. Association for Computing Machinery, New
  York, NY, USA, 95–101.
\newblock
\showISBNx{9781450359689}
\showDOI{%
\url{http://dx.doi.org/10.1145/3270316.3270611}}


\bibitem{triglianos2017measuring}
{Vasileios Triglianos}, {Sambit Praharaj}, {Cesare Pautasso}, {Alessandro
  Bozzon}, {and} {Claudia Hauff}. 2017.
\newblock \showarticletitle{Measuring student behaviour dynamics in a large
  interactive classroom setting}. In {\em Proceedings of the 25th Conference on
  User Modeling, Adaptation and Personalization}. 212--220.
\newblock


\bibitem{ALStrategies200}
{USF}. 2020.
\newblock Interactive Techniques for Teaching.
\newblock   (june 2020).
\newblock
\showURL{%
Retrieved June 22, 2020 from
  \url{https://www.usf.edu/atle/teaching/interactive-techniques.aspx}}


\bibitem{10.1145/3290605.3300548}
{Laton Vermette}, {Joanna McGrenere}, {Colin Birge}, {Adam Kelly}, {and}
  {Parmit~K. Chilana}. 2019.
\newblock \showarticletitle{Freedom to Personalize My Digital Classroom:
  Understanding Teachers’ Practices and Motivations}. In {\em Proceedings of
  the 2019 CHI Conference on Human Factors in Computing Systems} {\em (CHI
  ’19)}. Association for Computing Machinery, New York, NY, USA, 1–14.
\newblock
\showISBNx{9781450359702}
\showDOI{%
\url{http://dx.doi.org/10.1145/3290605.3300548}}


\bibitem{virtue2018dispositional}
{David~C Virtue}. 2018.
\newblock \showarticletitle{Dispositional, relational, and practical aspects of
  professional learning for middle level educators}.
\newblock {\em Preparing middle level educators for 21st century schools:
  Enduring beliefs, changing times, evolving practices\/} (2018), 365--375.
\newblock


\bibitem{walker2012course}
{Henry~M Walker}. 2012.
\newblock \showarticletitle{Course planning: the day-by-day course schedule}.
\newblock {\em ACM Inroads\/} {3}, 3 (2012), 22--24.
\newblock


\bibitem{weber2020efficient}
{Tobias Weber}. 2020.
\newblock \showarticletitle{Efficient strategies in course planning for
  low-resource minority language classes in higher education: observations from
  Uralic studies and the example of South Estonian}.
\newblock {\em The Language Learning Journal\/} (2020), 1--15.
\newblock


\bibitem{weiser2018medium}
{Orli Weiser}, {Ina Blau}, {and} {Yoram Eshet-Alkalai}. 2018.
\newblock \showarticletitle{How do medium naturalness, teaching-learning
  interactions and Students' personality traits affect participation in
  synchronous E-learning?}
\newblock {\em The internet and higher education\/}  {37} (2018), 40--51.
\newblock


\bibitem{weston1986selecting}
{Cynthia Weston} {and} {Patricia~A Cranton}. 1986.
\newblock \showarticletitle{Selecting instructional strategies}.
\newblock {\em The Journal of Higher Education\/} {57}, 3 (1986), 259--288.
\newblock


\bibitem{willson2019analysing}
{Rebekah Willson}. 2019.
\newblock \showarticletitle{Analysing qualitative data: you asked them, now
  what to do with what they said}. In {\em Proceedings of the 2019 conference
  on human information interaction and retrieval}. 385--387.
\newblock


\bibitem{10.1145/3313831.3376781}
{Rainer Winkler}, {Sebastian Hobert}, {Antti Salovaara}, {Matthias
  S\"{o}llner}, {and} {Jan~Marco Leimeister}. 2020.
\newblock \showarticletitle{Sara, the Lecturer: Improving Learning in Online
  Education with a Scaffolding-Based Conversational Agent}. In {\em Proceedings
  of the 2020 CHI Conference on Human Factors in Computing Systems} {\em (CHI
  ’20)}. Association for Computing Machinery, New York, NY, USA, 1–14.
\newblock
\showISBNx{9781450367080}
\showDOI{%
\url{http://dx.doi.org/10.1145/3313831.3376781}}


\bibitem{yardi2008whispers}
{Sarita Yardi}. 2008.
\newblock {\em Whispers in the Classroom}.
\newblock MacArthur Foundation Digital Media and Learning Initiative.
\newblock


\bibitem{yuliati2018influence}
{Lia Yuliati} {and} {Nuril Munfaridah}. 2018.
\newblock \showarticletitle{The Influence of Thinking Maps on Discovery
  Learning toward Physics Problem Solving Skills}. In {\em Proceedings of the
  2nd International Conference on Education and Multimedia Technology}. 59--63.
\newblock


\bibitem{10.1145/2212776.2212810}
{Panagiotis Zaharias}, {Marios Belk}, {and} {George Samaras}. 2012.
\newblock \showarticletitle{Employing Virtual Worlds for HCI Education: A
  Problem-Based Learning Approach}. In {\em CHI '12 Extended Abstracts on Human
  Factors in Computing Systems} {\em (CHI EA '12)}. Association for Computing
  Machinery, New York, NY, USA, 317–326.
\newblock
\showISBNx{9781450310161}
\showDOI{%
\url{http://dx.doi.org/10.1145/2212776.2212810}}


\end{thebibliography}

\end{document}